\documentstyle[12pt,a41,epsfig,cite,portland,rotating]{article}

\newcommand{\Mvec}{\mbox{\rm\bf M}}

\newcommand{\beq}{\begin{equation}}
\newcommand{\eeq}{\end{equation}}
\newcommand{\bea}{\begin{eqnarray}}
\newcommand{\eea}{\end{eqnarray}}

\newcommand{\lsim}{\raisebox{-0.07cm}{$\, \stackrel{<}{{\scriptstyle
\sim}}\, $}}

\newcommand\MeV{\,\mbox{MeV}}
\newcommand\GeV{\,\mbox{GeV}}

\newcommand{\tx}{\textrm}
\newcounter{lin}

\graphicspath{{./},
	      {./figures/}}

\begin{document}
\begin{titlepage}

\begin{flushleft}
DESY 05--012 \hfill {\tt hep-ph/0607200} \\
SFB-CPP-06/33 \\
%\today \\
\end{flushleft}

\vspace{3cm}
\noindent
\begin{center}
{\LARGE\bf Non--Singlet QCD Analysis of Deep Inelastic} \\
\vspace*{2mm}
\noindent
{\LARGE \bf World Data at \boldmath $O(\alpha_s^3)$ }
\end{center}
\begin{center}

\vspace{2.0cm}
{\large Johannes Bl\"umlein$^a$, Helmut B\"ottcher$^{a,b}$, and Alberto 
Guffanti$^a$\footnote{Present address: School of Physics,
University of Edinburgh (SUPA)
King's Buildings,
Mayfield Road,
Edinburgh EH9 3JZ, United Kingdom
}}

\vspace{1.5cm}
{\it 
$^a$Deutsches Elektronen Synchrotron, DESY}\\

\vspace{3mm}
{\it  Platanenallee 6, D--15738 Zeuthen, Germany}\\

\vspace{5mm}
{\it 
$^b$Institut f\"ur Physik, Humboldt--Universit\"at, Newtonstra\ss{}e. 15, D--12489 
Berlin} \\
%%\today
\vspace{3cm}
\end{center}
\begin{abstract}
\noindent
A non--singlet QCD analysis of the structure function $F_2(x,q^2)$ in LO, NLO, NNLO and 
N$^3$LO is performed based on the world data for charged lepton scattering. We 
determine the valence quark parton densities and present their parameterization 
and that of the correlated errors in a wide range of $x$ and $Q^2$. In the analysis we 
determined the QCD--scale $\Lambda_{\rm QCD, N_f = 4}^{\overline{\rm MS}} = 265 \pm 27 
\MeV~~{\rm (NLO)},~~226 \pm 25 \MeV ~~{\rm (NNLO)},~~234 \pm 26 \MeV~~{\rm (N^3LO)}$, 
with a remainder uncertainty of $\pm 2\MeV$ for the yet unknown 4--loop anomalous dimension,
corresponding to $\alpha_s(M_Z^2) = 0.1148 \pm 0.0019~~{\rm NLO},~~0.1134 {\tiny{\begin{array}{c} 
+0.0019 \\ -0.0021 \end{array}}}~~{\rm NNLO},~~0.1141 {\tiny{\begin{array}{c} +0.0020 \\ 
-0.0022 \end{array}}}~~{\rm N^3LO}$. A comparison is performed to other determinations
of the QCD scale and $\alpha_s(M_Z^2)$ in deeply inelastic scattering. The higher twist 
contributions of $F_2^p(x,Q^2)$ and $F_2^d(x,Q^2)$ are extracted in the large $x$ region,
subtracting the twist--2 contributions obtained in the NLO, NNLO and N$^3$LO analysis.
\end{abstract}

\end{titlepage}

\newpage
\sloppy

%%%%%%%%%%%%%%%%%%%%%%%%%%%%%%%%%%%%%%%%%%%%%%%%%%%%%%%%%%%%%%%%%%%%%%%
%	Introduction
%%%%%%%%%%%%%%%%%%%%%%%%%%%%%%%%%%%%%%%%%%%%%%%%%%%%%%%%%%%%%%%%%%%%%%%
\section{Introduction}
\label{sec:introduction}
%%%%%%%%%%%%%%%%%%%%%%%%%%%%%%%%%%%%%%%%%%%%%%%%%%%%%%%%%%%%%%%%%%%%%%%

\vspace{1mm}\noindent
Deeply inelastic scattering (DIS) at large 4--momentum transfer $Q^2  = - q^2$ 
offers the possibility to probe the partonic sub--structure of the nucleon.
The light--cone expansion \cite{LC} allows to describe the forward Compton
amplitude in terms of the twist--expansion. For the twist--2 contributions
the factorization theorems hold and the calculation of the scaling 
violations of the parton distribution functions and structure functions 
is possible within the framework of perturbative Quantum Chromodynamics 
(QCD). In the present paper we determine the flavor non--singlet parton 
distribution functions $xu_v(x,Q^2)$ and $xd_v(x,Q^2)$ using the available 
$e(\mu) p$ and  $e(\mu) d$ world data \cite{BCDMS,SLAC,NMC,H1,ZEUS} up to  
the next-to-next-to leading (NNLO) and next-to-next-to-next to leading 
order level (N$^3$LO). This is possible due to the advent of the 3--loop 
non--singlet anomalous dimensions \cite{Moch:2004pa} and 3--loop Wilson 
coefficients \cite{Vermaseren:2005qc}\footnote{Moments of the 3--loop 
anomalous dimension and Wilson coefficients were calculated in \cite{MOM}.}, 
as well as first informations on the 4--loop non--singlet anomalous 
dimension, with its second moment \cite{BC}. The QCD analysis of the world 
non--singlet data has the advantage to deal with a restricted set of parton 
distribution functions, to which the virtual photon couples directly and 
which do therefore contribute already at leading order. The analysis is, 
furthermore, free of assumptions on the gluon distribution function which 
stabilizes the systematics since  a lower number of parameters describing the 
parton distributions has to be measured and the analysis is free of the 
correlation between the  QCD scale $\Lambda_{\rm QCD}$ and the gluon density. 
Also due to the particular behaviour of the 
flavor non--singlet parton densities in the smaller Bjorken--$x$ region 
in comparison to the sea--quark and gluon densities a separate non--singlet 
analysis is in order to better understand systematics effects present in 
combined non--singlet and singlet analyses.

Previous 3--loop QCD analyses were mainly performed as combined singlet 
and non--singlet analyses \cite{MRST03,A02}\footnote{A recent update of 
the latter analysis has been given in \cite{Alekhin:2006zm} very recently.}, 
partly based on preliminary, approximative expression of the 3--loop 
splitting functions. 
Other analyses were carried out for fixed moments only in the singlet 
\cite{SY,KAT} and non-singlet case analyzing neutrino data 
\cite{SY}\footnote{For earlier results see \cite{SYS}.}. First 
results of our non--singlet analysis were published in \cite{BBG04}. Very 
recently a 3--loop non--singlet analysis was also carried out in 
Ref.~\cite{Gluck:2006yz}. There are numerous NLO analyses, among them 
\cite{BCDMS,H1,ZEUS,CTEQ_P}, which we will use in comparisons below. 
Another recent NLO parameterization \cite{Pumplin:2005rh} fixes $\alpha_s(M_Z^2)$
and presents a series of ten parton distribution sets associated to equidistant
values of the strong coupling constant in the range  
$\alpha_s(M_Z^2)~\epsilon~[0.110,~0.128]$.

The non-singlet analysis will provide us with the distribution functions
$xu_v(x,Q^2)$ and $xd_v(x,Q^2)$ with correlated errors. From these 
distributions we may form Mellin--moments which can be compared to 
lattice--simulations of these quantities. The lattice calculations have to 
be performed at as low as possible pion masses due to a potential 
non--linear behaviour on $m_{\pi}$ in the small mass region and  
non--perturbative renormalization has to be used. Precision results of these 
simulations with $N_f =2$ dynamical quarks are expected soon. The comparison 
of the Mellin moments obtained in a 3(4)--loop QCD analysis of the non--singlet 
World Data at the one side and lattice simulations on the other side
will lead to new important tests of Quantum Chromodynamics.

The paper is organized as follows. In Section~2 we briefly summarize 
the basic formalism. The data analysis and calculation of errors is 
outlined in Section~3. Different corrections such as the account for 
a finite sea quark difference $x (\overline{d} - \overline{u})(x,Q^2)$, 
heavy quark contributions, target mass and deuteron wave function corrections 
are dealt with in  Section~4. The fit results for the non--singlet parton 
distribution functions, their correlated errors and evolution are presented 
in Section~5, where also comparisons to other analyses are made. In Section~6 the 
results on $\Lambda_{\rm QCD}$ and $\alpha_s(M_Z^2)$ in NLO, NNLO and N$^3$LO are 
given and compared to other NLO and NNLO results. In Section~7 we present
our results on the higher twist contributions in the large $x$ region for
the structure functions $F_2^p(x,Q^2)$ and $F_2^d(x,Q^2)$ and 
Section 8 contains the conclusions.
%%%%%%%%%%%%%%%%%%%%%%%%%%%%%%%%%%%%%%%%%%%%%%%%%%%%%%%%%%%%%%%%%%%%%%%
%	Formalism
%%%%%%%%%%%%%%%%%%%%%%%%%%%%%%%%%%%%%%%%%%%%%%%%%%%%%%%%%%%%%%%%%%%%%%%
\section{Basic Formalism}
\label{sec:formalism}
%%%%%%%%%%%%%%%%%%%%%%%%%%%%%%%%%%%%%%%%%%%%%%%%%%%%%%%%%%%%%%%%%%%%%%%

\vspace{1mm}\noindent
We choose the following parameterization for the valence quark densities
%-----------------------------------------------------------------------
{\small
\begin{eqnarray}
\label{equ:param}
x v_i(x,Q^2) = A_i x^{a_i}(1-x)^{b_i} (1 + \rho_i \sqrt{x} + \gamma_i x)~,
\end{eqnarray}
}
%-----------------------------------------------------------------------

\noindent
with $v_i = u_v, d_v,~i=u,d$. Three valence--projections are considered.
In the region $x \leq 0.3$ for the difference of the proton and deuteron 
data we use
%-----------------------------------------------------------------------
\begin{eqnarray}
f^{\rm NS}(x,Q^2) = \frac{1}{3}\left[ x(u + \overline{u})(x,Q^2) - 
x(d + \overline{d})(x,Q^2)\right]~.
\end{eqnarray}
%-----------------------------------------------------------------------
The combinations of parton densities in the valence region $x > 0.3$ for 
$F_2^{p,d}(x,Q^2)$ are
%-----------------------------------------------------------------------
\begin{eqnarray}
f^v_p(x,Q^2) &=& \frac{5}{18}\left[ x(u - \overline{u})(x,Q^2) + 
x(d - \overline{d})(x,Q^2)\right] 
+ \frac{1}{6}\left[ x(u - \overline{u})(x,Q^2) 
- x(d - \overline{d})(x,Q^2)\right]  
\nonumber\\ 
\\
f^v_d(x,Q^2) &=& \frac{5}{18}\left[ x(u - \overline{u})(x,Q^2) + 
x(d - \overline{d})(x,Q^2)\right]~. 
\end{eqnarray}
%-----------------------------------------------------------------------
The evolution equations are solved in Mellin-$N$ space and the Mellin 
transforms of the above distributions are denoted by $f^{\rm NS}(N,Q^2),
f^v_{p,d}(N,Q^2)$, respectively. The non--singlet structure functions  
are given by
%-----------------------------------------------------------------------
\begin{eqnarray}
F_k(N,Q^2) = \left[1 + a_s(Q^2) C_1(N) + a_s^2(Q^2) C_2(N) + 
a_s^3(Q^2) C_3(N)\right]
             f_k(N,Q^2)
\end{eqnarray}
%-----------------------------------------------------------------------
for the three cases above. Here $a_s(Q^2) = \alpha_s(Q^2)/(4\pi)$ denotes
the strong coupling constant and $C_i(N)(Q^2)$ are the non--singlet
Wilson coefficients in $O(a_s^i)$ \cite{FP,NS2,Vermaseren:2005qc}. 
For the representation of the Wilson coefficients we used
those given in \cite{vanNeerven:1999ca,Vermaseren:2005qc} which are based 
on numerical fits. For the 2--loop Wilson coefficients and 3--loop anomalous 
dimensions compact analytic representations are also available, based on the 
reduction of the respective harmonic sums to 5 (14) basic functions
\cite{BK1, ANCONT, JB1, BM1, BM2}.\footnote{Similar representations were  
derived for the 2--loop Wilson coefficients for the unpolarized and 
polarized Drell--Yan process and scalar and pseudoscalar Higgs--boson 
production at hadron colliders in the heavy mass limit as well as for the
hadron fragmentation functions in $e^+e^-$ annihilation \cite{BR1,BR2}.}

The solution of the non--singlet evolution equation for the parton 
densities to 4--loop order reads
%-----------------------------------------------------------------------
\begin{eqnarray}
F_k(N,Q^2) &=& F_k(N,Q_0^2) 
\left(\frac{a}{a_0}\right)^{-\hat{P}_0(N)/{\beta_0}}
\Biggl\{1 - \frac{1}{\beta_0} (a - a_0) \left[\hat{P}_1^+(N)
- \frac{\beta_1}{\beta_0} \hat{P}_0 \right] \nonumber\\ 
& & - \frac{1}{2 \beta_0}\left(a^2 - a_0^2\right) \left[\hat{P}_2^+(N) 
- \frac{\beta_1}{\beta_0} \hat{P}_1^+ + \left( \frac{\beta_1^2}{\beta_0^2}
- \frac{\beta_2}{\beta_0} \right) \hat{P}_0(N)   \right]
\nonumber\\ & &
+ \frac{1}{2 \beta_0^2} \left(a - a_0\right)^2 \left(\hat{P}_1^+(N)
- \frac{\beta_1}{\beta_0} \hat{P}_0 \right)^2
\nonumber\\ & &
%---
- \frac{1}{3 \beta_0} \left(a^3 - a_0^3\right)
\Biggl[\hat{P}_3^+(N)
- \frac{\beta_1}{\beta_0} \hat{P}_2^+(N)  
+ \left(\frac{\beta_1^2}{\beta_0^2} - 
\frac{\beta_2}{\beta_0}\right) \hat{P}_1^+(N)
\nonumber\\ & & 
+\left(\frac{\beta_1^3}{\beta_0^3} -2 \frac{\beta_1 \beta_2}{\beta_0^2} + 
\frac{\beta_3}{\beta_0} \right) \hat{P}_0(N)  \Biggr] 
\nonumber\\ & &
+
\frac{1}{2 \beta_0^2} \left(a-a_0\right)\left(a_0^2 - a^2\right)
\left(\hat{P}_1^+(N)-\frac{\beta_1}{\beta_0} \hat{P}_0(N) \right) 
\nonumber\\ & & \times
\left[\hat{P}_2(N) - \frac{\beta_1}{\beta_0} \hat{P}_1(N) - 
\left(\frac{\beta_1^2}{\beta_0^2} 
- \frac{\beta_2}{\beta_0} \right) \hat{P}_0(N) 
 \right]
\nonumber\\ & &
- \frac{1}{6 \beta_0^3} \left(a-a_0\right)^3 
\left(\hat{P}_1^+(N)-\frac{\beta_1}{\beta_0} \hat{P}_0(N)  
\right)^3 \Bigg\}~.
\end{eqnarray}
%-----------------------------------------------------------------------
Here, $\hat{P}_k$ denote the Mellin transforms of the $(k+1)$--loop splitting 
functions.

The strong coupling constant obeys the evolution equation 
%-----------------------------------------------------------------------
\begin{eqnarray}
\label{eqAS}
\frac{da_s(Q^2)}{d \ln Q^2} = - \sum_{k=0}^{\infty} \beta_k 
a_s^{k+2}(Q^2)~.
\end{eqnarray}
%-----------------------------------------------------------------------
Eq.~(\ref{eqAS}) is solved in the $\overline{\rm MS}$--scheme applying 
the matching of flavor thresholds at $Q^2 = m_c^2$ and $Q^2 = m_b^2$ with
$m_c = 1.5 \GeV$ and $m_b = 4.5 \GeV$ as described in \cite{CHET,SB}.
The convention for the $\overline{\rm MS}$--scheme introduced in \cite{BAR}
is extended in this way. To be capable to compare with other measurements
of $\Lambda_{\rm QCD}$ we adopt this prescription.
%%%%%%%%%%%%%%%%%%%%%%%%%%%%%%%%%%%%%%%%%%%%%%%%%%%%%%%%%%%%%%%%%%%%%%%
\section{Data Analysis and Error Calculation}
\label{sec:data}
%%%%%%%%%%%%%%%%%%%%%%%%%%%%%%%%%%%%%%%%%%%%%%%%%%%%%%%%%%%%%%%%%%%%%%%

\vspace{1mm}
\noindent
For the non--singlet QCD analysis presented in the present paper  
structure function data measured in charged lepton proton and  deuteron 
deep inelastic scattering are used. The experiments contributing to
the statistics are BCDMS~\cite{BCDMS}, SLAC~\cite{SLAC}, NMC~\cite{NMC}, 
H1~\cite{H1}, and ZEUS~\cite{ZEUS}. The BCDMS data were recalculated 
replacing $R_{QCD}$ by $R_{1998}$ \cite{R1998}. The SLAC data come from a 
re-analysis of the data from the experiments E49, E61, E87, and E89. All 
deuteron data were corrected for Fermi motion and off-shell
effects~\cite{MT}. We form three data samples~: $F_2^p(x,Q^2)$, 
$F_2^d(x,Q^2)$ in the valence quark region $x \geq 0.3$ and $F_2^{NS} = 2
(F_2^p - F_2^d)$ in the region $x < 0.3$. In the valence quark region we 
approximate the parton distribution functions by pure valence quarks. 
Only data with $Q^2 > 4~\GeV^2$ were included in the analysis and a cut in the hadronic mass 
of $W^2 > 12.5~\GeV^2$ was applied in order to widely eliminate  higher 
twist effects from the data samples. After these cuts we are left with 762 
data points, 322 for $F_2^p$, 232 for $F_2^d$, and 208 for $F_2^{NS}$. The 
BCDMS data have been used with an additional cut of $y_{\mu} > 0.3$ in 
order to exclude a region with significant correlated systematic
errors, cf. also~\cite{H1}. In case of $e(\mu)N$ deep inelastic scattering 
experiments measuring the 
kinematic variables either only form the lepton or the outgoing 
jet--kinematics the systematics of the respective measurement becomes 
worse at low values of $y$ (lepton measurement) or high values of $y$ (jet 
measurement). These effects were studied in Ref.~\cite{JBMK} and give rise
to the $y$--cut mentioned for the BCDMS-data.
A cut of $Q^2 > 8~{\rm \GeV^2}$ was imposed on the
NMC data. These cuts reduce the total number of data points available
for the analysis from 762 to 551. 

The data sets used contain both statistical and systematic errors which in 
turn, for some data sets, split into uncorrelated and correlated ones. 
There are several procedures proposed to treat the experimental errors 
and the determination of uncertainties from parton distributions~\cite{dPDF}.  
Here we decided to use the simplest procedure by adding the statistical 
and systematic errors, uncorrelated and correlated ones, in quadrature 
when using the data sets in the fit. However, we allowed for a relative 
normalization shift $N_i$ between the different data sets within the normalization 
uncertainties $\Delta N_i$ quoted by the experiments or assumed accordingly. These 
normalization shifts were fitted once and then kept fixed. The number of 
data points, their $x$ and $Q^2$ range, and the normalization shifts 
determined are summarized in Table~1. 

The normalization shift for each data set enters as an additional term in 
the $\chi^2$--expression which then reads
%------------------------------------------------------------------------
\begin{equation}
\chi^2 = \sum_{i=1}^{n^{exp}} \left [ 
         \frac {(N_i - 1)^2}
               {(\Delta N_i)^2} + 
         \sum_{j=1}^{n^{data}} 
         \frac {(N_i F_{2,j}^{data} - F_{2,j}^{theor})^2}
               {(N_i\Delta F_{2,j}^{data})^2} 
         \right ],
\end{equation}
%------------------------------------------------------------------------

\noindent
where the sums run over all data sets and in each data set over all data 
points. The minimization of the $\chi^2$ value above to determine the best 
parameterization of the unpolarized parton distributions is done using the 
program {\tt MINUIT} \cite{MINUIT}. Only fits giving a positive definite
covariance matrix at the end have been accepted in order to be able to 
calculate the fully correlated $1\sigma$ error bands.

\vspace{2mm}
\noindent
The  $1\sigma$ error for the parton density $f_q$ as
given by Gaussian error propagation is

%------------------------------------------------------------------------
\beq
\sigma( f_q(x))^2 =  \sum_{i,j=1}^{n_p} 
                \left( \frac{\partial f_q}{\partial p_i}
                \frac{\partial f_q}{\partial p_j} \right)
                \tx{cov}(p_i,p_j)~, 
\eeq
%------------------------------------------------------------------------

\noindent
where the sum runs over all fitted parameters. The functions $\partial f_q /
\partial p_i$ are the derivatives of $f_q$ w.r.t. the fit parameter $p_i$,
and $\tx{cov}(p_i,p_j)$ are the elements of the covariance matrix. 
The derivatives $\partial f_q / \partial p_i$ can be calculated analytically 
at the input scale $Q_0^2$. Their values at $Q^2$ are given by evolution 
which in our case is performed in {Mellin--$N$} space. The evolution 
formalism in {Mellin--$N$} space is such that the input and the evolution 
parts factorize which leads to a straightforward analytic error calculation. The 
derivatives evolved in {Mellin--$N$} space are transformed back to 
$x$--space and can then be used according to the formula above together 
with the elements of a positive definite covariance matrix determined at 
the input scale $Q_0^2$ by the QCD fit to the data. 

Let us discuss the derivatives in {Mellin--$N$} space a bit further.
The {Mellin--$N$} moment for complex values of $N$ calculated at the
input scale $Q_0^2$ for the parton density parameterized as in 
Eq.~(\ref{equ:param}) is given by 

%------------------------------------------------------------------------
\bea
\Mvec[f(x,a,b,\rho,\gamma)](N) & = & A \left[B(a+N+1,b+1) + \rho
B(a+N-1/2,b+1) \right. \nonumber \\  
                               &   & + \left. \gamma B(a+N,b+1)\right],
\eea
%------------------------------------------------------------------------

\noindent
with the normalization constant

%------------------------------------------------------------------------
\beq
A = \frac{C}{\left[B(a,b+1) + \rho B(a+1/2,b+1) + \gamma 
B(a+1,b+1)\right]}~.
\eeq
%------------------------------------------------------------------------

\noindent
Here $C$ is the respective number of valence quarks, i.e. $C_{u_v} =
2$ and $C_{d_v} = 1$, and $B(a,b) = \Gamma(a)\Gamma(b)/\Gamma(a+b)$
denotes the Euler Beta--function for complex arguments.

The general form of the derivative of the Mellin moment
$\Mvec$ w.r.t. the parameter $p$ is given by  

%------------------------------------------------------------------------
\beq
\frac {\partial {\Mvec}[f(x,p)](N)}{\partial p} = A
\frac {\partial \overline {\Mvec}} {\partial p} + 
\overline {\Mvec} \frac {\partial A}{\partial p},
\eeq
%------------------------------------------------------------------------

\noindent
with $\overline {\Mvec} = {\Mvec} / A$. In the present analysis only
the parameters $a$ and $b$ have been fitted for both the $u_v$ and the
$d_v$ parton distribution while the other parameters involved are kept
fixed after a first minimization since their errors turned out to be rather large 
compared to the central values. For the parton densities
$u_v(x,Q_0^2)$ and $d_v(x,Q_0^2)$ the derivatives w.r.t. parameter $a$ are: 

%------------------------------------------------------------------------
\bea
\frac {\partial{\overline{\Mvec}}}{\partial a} & = & \bigl\{ 
[\Psi(a-1+N) - \Psi(a+N+b)] + \gamma\frac{a-1+N}{a+N+b} (\Psi(a+N) -
\bigr. \nonumber \\ 
  & & \Psi(a+N+b+1)) \bigr\} B(a-1+N,b+1) + \rho
[\Psi\left(a-\frac{1}{2}+N\right) - \nonumber \\  
  & & \Psi\left(a+\frac{1}{2}+N+b\right)] 
B\left(a-\frac{1}{2}+N,b+1\right)\;
\eea
%------------------------------------------------------------------------

\noindent
and

%------------------------------------------------------------------------
\bea
\hspace*{-6.0cm}
\frac {\partial A}{\partial a} & = & - A Z_a / X_a = - C Z_a / X_a^2\;,  
\eea
%------------------------------------------------------------------------

\noindent
where

%------------------------------------------------------------------------
\bea
Z_a  & = & [\Psi(a) - \Psi(a+b+1)] B(a,b+1) + \rho 
\left[\Psi\left(a+\frac{1}{2}\right) -
\Psi\left(a+\frac{1}{2}+b+1\right)\right] \nonumber \\ 
& & B(a+\frac{1}{2},b+1) + \gamma [\Psi(a+1) - \Psi(a+1+b+1)]
B(a+1,b+1)\;, 
\eea
%------------------------------------------------------------------------

\noindent
and

%------------------------------------------------------------------------
\bea
\hspace*{-3.50cm}
X_a & = & B(a,b+1) + \rho B\left(a+\frac{1}{2},b+1\right) + \gamma 
B(a+1,b+1).
\eea
%------------------------------------------------------------------------

\noindent
For the derivatives of $u_v(x,Q_0^2)$ and $d_v(x,Q_0^2)$ w.r.t. parameter $b$
one obtains:   

%------------------------------------------------------------------------
\bea
\frac {\partial{\overline{\Mvec}}}{\partial b} & = & \bigl\{ 
[\Psi(b+1) - \Psi(a+N+b)] + \gamma\frac{a-1+N}{a+N+b} (\Psi(b+1) -
\bigr. \nonumber \\ 
  & & \Psi(a+N+b+1)) \bigr\} B(a-1+N,b+1) + \rho
\left[\Psi(b+1) - \nonumber \right.\\  
  & & \left. \Psi\left(a+\frac{1}{2}+N+b\right)\right] 
B\left(a-\frac{1}{2}+N,b+1\right)\;
\eea
%------------------------------------------------------------------------

\noindent
and

%------------------------------------------------------------------------
\bea
\hspace*{-6.0cm}
\frac {\partial A}{\partial b} & = & - A Z_b / X_b = - C Z_b / X_b^2\;,  
\eea
%------------------------------------------------------------------------

\noindent
where

%------------------------------------------------------------------------
\bea
Z_b  & = & [\Psi(b+1) - \Psi(a+b+1)] B(a,b+1) + \rho [\Psi(b+1) -
\Psi(a+\frac{1}{2}+b+1)] \nonumber \\ 
& & B(a+\frac{1}{2},b+1) + \gamma [\Psi(b+1) - \Psi(a+1+b+1)]
B(a+1,b+1)\;, 
\eea
%------------------------------------------------------------------------

\noindent
and

%------------------------------------------------------------------------
\bea
\hspace*{-3.50cm}
X_b & = & B(a,b+1) + \rho B(a+\frac{1}{2},b+1) + \gamma B(a+1,b+1).
\eea
%------------------------------------------------------------------------

\noindent
Here $\Psi(z) = d \ln \Gamma(z)/dz$ is Euler's $\Psi$--function.
To obtain the gradients
for the error calculation of the structure functions $F_2^p$, $F_2^d$,
and $F_2^{\rm NS}$ the relevant gradients of the parton distribution functions 
in Mellin space have to be multiplied with the
corresponding Wilson coefficients.

This yields the errors as far as the QCD parameter $\Lambda_{\rm QCD}$ is
fixed and regarded uncorrelated. The error calculation for a
variable $\Lambda_{\rm QCD}$ is done numerically due to the non--linear 
relation and required iterative treatment in the calculation of
$\alpha_s(Q^2,\Lambda_{\rm QCD})$. The respective gradients are given
by performing the evolution for $\Lambda \pm \delta$, with $\delta
\ll \Lambda$, and calculating

%------------------------------------------------------------------------
\beq
\frac {\partial f(x,Q^2,\Lambda)}{\partial \Lambda} \simeq 
       \frac {f(x,Q^2,\Lambda + \delta)-f(x,Q^2,\Lambda -
\delta)}{2\delta}  
\eeq
%------------------------------------------------------------------------

\noindent
using typical values for $\delta \lsim  10~\MeV$ in the present analysis.
%%%%%%%%%%%%%%%%%%%%%%%%%%%%%%%%%%%%%%%%%%%%%%%%%%%%%%%%%%%%%%%%%%%%%%%
%	Corrections
%%%%%%%%%%%%%%%%%%%%%%%%%%%%%%%%%%%%%%%%%%%%%%%%%%%%%%%%%%%%%%%%%%%%%%%
\section{Corrections}
\label{sec:corrections}
%%%%%%%%%%%%%%%%%%%%%%%%%%%%%%%%%%%%%%%%%%%%%%%%%%%%%%%%%%%%%%%%%%%%%%%

\vspace{1mm}\noindent
To have the possibility to include deuteron data into the analysis 
deuteron wave function corrections both for Fermi motion and 
off-shell effects~\cite{MT} have to be carried out. These corrections 
are performed  during the fit.

Since we analyze charged lepton--nucleon scattering the non--singlet 
structure functions require the distribution $x(\overline{d} - 
\overline{u})(x,Q^2)$ as an input, which cannot easily be extracted in the 
DIS analysis, but can be determined from Drell--Yan data \cite{E866}. We
refer to a parameterization in \cite{MRST02} at $Q^2 = 1 \GeV^2$ which has 
been evolved to $Q_0^2 = 4 \GeV^2$. An illustration is given in Figure~1, 
containing also the older parton parameterization \cite{A02} which has 
been updated recently \cite{Alekhin:2006zm}. 

Part of the data is situated at low values of $Q^2$ and large values of $x$
which have to be corrected for target mass effects, cf. \cite{GP76}. The 
respective numerical effects are discussed below, see Figures~5, 6 and 8.

Beginning with $O(\alpha_s^2)$ the non--singlet structure functions 
receive heavy flavor contributions due to the Compton--process
$\gamma^* + q(\overline{q}) \rightarrow q(\overline{q}) + g^* \rightarrow 
Q\overline{Q}$ and 
$\gamma^* + g^* \rightarrow Q\overline{Q}$, where the off--shell gluon is 
emitted from a light quark or anti--quark line in lowest order, 
\cite{HF1}. As the present analysis is carried out in Mellin space one 
needs accurate and fast representations for the heavy flavor contributions, which were
derived in Ref.~\cite{Alekhin:2003ev} 
for all available heavy flavor Wilson coefficients. 
We accounted for these contributions in the present analysis.
The numerical effects 
of these contributions for the case of $c\overline{c}$ production are 
shown in Figures~2, 3 and 4 for the proton and 
deuteron valence regions $x \geq 0.3$ and for $F_2^{\rm NS}(x,Q^2)$ at 
$x \leq 0.3$ for a wide region in $Q^2$. The respective arrows mark the 
lowest data point in the analysis. Except of the very high $Q^2$ region, 
$Q^2 \simeq 10000 \GeV^2$, where the relative contribution amounts to 
$O(0.4\%)$ the typical corrections are (far) below 1 per mille and are 
therefore widely negligible. The present analysis would require  
the $O(\alpha_s^3)$ heavy flavor corrections for $F_2^{\rm NS}(x,Q^2)$ as well, 
which were not yet calculated.\footnote{At $O(\alpha_s^3)$ only the corrections for
$F_L(x,Q^2)$ in the region $Q^2 \gg m_Q^2$ are known, 
cf. \cite{HF2,HF3}.} From the $O(\alpha_s^2)$ results we do not expect 
large additional effects in the non--singlet case, however. The 
$b\overline{b}$--contributions are even smaller.

Universal small-$x$ contribution to the evolution kernel for deep--inelastic 
scattering were derived in \cite{Kirschner:1983di}
 in leading order. Resumming these
effects with renormalization group techniques yields contributions on the
level of 1--2\% and less depending on the size of the subleading terms
\cite{BV}, which are important. Like in the flavor singlet case \cite{BV1} 
the sub-, subsub-leading 
contributions, etc. turn out to be of quite similar size, however, with alternating sign,
as the respective next term. This is easily verified expanding the available fixed oder 
results in consecutive powers in Mellin $1/N^k$, cf. also \cite{Moch:2004pa}.
Hence, these effects do not affect the present analysis. 

%%%%%%%%%%%%%%%%%%%%%%%%%%%%%%%%%%%%%%%%%%%%%%%%%%%%%%%%%%%%%%%%%%%%%%%
%	Fit Results
%%%%%%%%%%%%%%%%%%%%%%%%%%%%%%%%%%%%%%%%%%%%%%%%%%%%%%%%%%%%%%%%%%%%%%%
\section{Fit Results}
\label{sec:fitres}
%%%%%%%%%%%%%%%%%%%%%%%%%%%%%%%%%%%%%%%%%%%%%%%%%%%%%%%%%%%%%%%%%%%%%%%

\vspace{1mm}\noindent
The flavor non--singlet analysis relies on the three complementary data sets:
the structure functions $F_2^{p,d}(x,Q^2)$ in the valence--quark region 
$x \geq 0.3$ and the combination of these structure functions
$F_2^{\rm NS}(x,Q^2) = 2 [F_2^p(x,Q^2) - F_2^d(x,Q^2)]$~. To isolate the twist--2 
contributions to the deep--inelastic structure functions
$F_{2}^{p,d}(x,Q^2)$ cut--studies in $Q^2$ and $W^2$ were performed. It turns out, that
power corrections are widely absent in the kinematic region
$Q^2 \geq 4 \GeV^2, W^2 > 12.5 \GeV^2$~.
We therefore choose these cuts for the whole data set to perform a twist--2 non--singlet
QCD analysis. 
We further unfold the target mass corrections \cite{GP76} for all data.

In Figure~5 the proton data for $F_2(x,Q^2)$ are shown in the valence quark region 
$x \geq 0.3$ indicating the above cuts by an arrow. The solid lines correspond to the 
NNLO QCD fit and the dashed lines correspond to the NNLO QCD fit adding target mass 
corrections. We extrapolate these fits to the region $12.5 \GeV^2 \geq W^2
\geq 4 \GeV^2$. There, at higher values of $x$ a clear gap between the data and 
the QCD fit is seen. Figure~6 shows the corresponding results for the deuteron data. 
In the kinematic region $x \leq 0.3$ the above $W^2$ cut is ineffective. Figure~7 
shows the result of the NNLO fit for this region. The effect of target mass corrections 
in the latter kinematic domain is shown in Figure~8, which turns out to 
be rather small.

In Table~2 we summarize the fit results for the parameters of the parton densities
$xu_v(x,Q^2_0)$, $xd_v(x,Q^2_0)$ and $\Lambda_{\rm QCD}^{\rm N_f =4}$ in the
$\overline{\rm MS}$--scheme in NLO, NNLO, and N$^3$LO. The value of $\chi^2/ndf$ 
at the minimum improves from 0.89 at NLO to 0.84 at N$^3$LO. In all the fits 
the covariance matrix is positive definite. As an example we show the covariance matrix 
of the NNLO fit in Table~3.   

Figure~9 illustrates our fit results for $xu_v(x,Q^2_0)$, $xd_v(x,Q^2_0)$ at
$Q_0^2 = 4 \GeV^2$ at NNLO with correlated errors (left figures). 
We compare to the results of \cite{MRST04} and a very recent analysis 
\cite{Alekhin:2006zm}. While there is very good agreement in the case of $xu_v(x,Q^2_0)$ some differences
at the $1\sigma$ level are visible for $xd_v(x,Q^2_0)$.
We also compare our results to some NLO analyses \cite{CTEQ_P,H1,GRV98} 
(right figures). The spread here turns out to be somewhat larger than in the
NNLO case. Note that the analyses we compare to are combined non-singlet/
singlet analyses and already different assumptions on the sea quark distributions
may cause deviations for the non--singlet distributions in the range of 
smaller values of $x$.

In Figure~10 (right figures) we show the convergence of our analysis
from LO to N$^3$LO, which demonstrates perturbative stability for the parton distribution 
functions.  
The lines rise in the large $x$ region and drop slightly for low values of $x$.
The shifts from NNLO to N$^3$LO are far within the present error bands.

Another way to compare the NNLO fit results consists in forming moments
of the distributions $u_v(x,Q^2), d_v(x,Q^2),$ and $u_v(x,Q^2)- d_v(x,Q^2)$.
These moments may be directly compared with the results of lattice simulations
at low pion masses using dynamical quarks in the near future.
In Table~4 we present the lowest non--trivial moments of these distributions
at $Q^2 = Q_0^2$ in N$^3$LO and NNLO and compare to the respective moments
obtained for the parameterizations \cite{MRST04,A02,Alekhin:2006zm}. 
The recent parameterization \cite{Alekhin:2006zm} comes somewhat closer 
to our present parameterization then an earlier one \cite{A02} due to an improvement
in the $x(\overline{d}(x,Q^2)-\overline{u}(x,Q^2))$ distribution and taking into account
off--shell effects in the target mass corrections \cite{PC}.
The errors of
the moments for \cite{Alekhin:2006zm} are slightly smaller than those in the
present analysis. There are still some  deviations, in particular for $d_v(x,Q^2)$, 
but most of the values get very close.  

In Figures~11 and 12 we show the evolution of the valence quark densities
$xu_v(x,Q^2)$ and $xd_v(x,Q^2)$ from $Q^2 = 1 \GeV^2$ to $Q^2 = 10^4 \GeV^2$
in the region $x~\epsilon~[10^{-3}, 1]$ at NNLO with correlated errors. 
To compute the respective error bands error propagation has to be performed
through the evolution equations. We also compare to other NNLO analyses 
\cite{MRST04, A02}. With rising values of $Q^2$ the distributions flatten 
at large values of $x$ and rise at low values. Numerically fast parameterizations 
of the distributions  $xu_v(x,Q^2)$ and $xd_v(x,Q^2)$ and their errors in the region
$Q^2 = [1,10^6] \GeV^2,~~x~\epsilon~[10^{-9},1]$ are performed and
are available from the authors on request.
%%%%%%%%%%%%%%%%%%%%%%%%%%%%%%%%%%%%%%%%%%%%%%%%%%%%%%%%%%%%%%%%%%%%%%%
%	alphas & lambda
%%%%%%%%%%%%%%%%%%%%%%%%%%%%%%%%%%%%%%%%%%%%%%%%%%%%%%%%%%%%%%%%%%%%%%%
\section{\boldmath \bf $\Lambda_{\rm QCD}$ and $\alpha_s(M_Z^2)$}
\label{sec:alplam}
%%%%%%%%%%%%%%%%%%%%%%%%%%%%%%%%%%%%%%%%%%%%%%%%%%%%%%%%%%%%%%%%%%%%%%%

\vspace{1mm}\noindent
The QCD analysis is performed in NLO, NNLO and N$^3$LO. The QCD scale 
$\Lambda_{\rm QCD}^{\rm N_f =4}$ is determined together with  the 
parameters of the parton distributions with correlated errors. As an 
example the correlation matrix elements of the NNLO analysis are shown in 
Table~3.

In spite of the unknown 4--loop anomalous dimensions one may wonder whether 
any statement can be made on the non--singlet parton distributions and 
$\Lambda_{\rm QCD}$ at the 4--loop level. The 3--loop Wilson coefficients
are known \cite{Vermaseren:2005qc} and the question arises, which effect 
has the 4--loop anomalous dimension if compared to the Wilson coefficient. 
The gross behaviour of the 4--loop anomalous dimension may be 
Pad\'{e}--approximated and we will allow a $\pm$ 100\% error for it in the 
analysis. After the present analysis was finished, the 2nd moment of the 
4--loop anomalous dimension was calculated in \cite{BC} 
%----------------------------------------------------------------------------
\begin{eqnarray}
\gamma_2 &=& 
\gamma_2^{(0)} a_s
+\gamma_2^{(1)} a_s^2
+\gamma_2^{(2)} a_s^3
+\gamma_2^{(3)} a_s^4 \nonumber\\ &=&
\frac{32}{9} a_s + \frac{9440}{243} a_s^2 
+ \left[\frac{3936832}{6561} - \frac{10240}{81} \zeta_3\right] a_s^3 
\nonumber\\ & &
+\left[\frac{1680283336}{1777147} - \frac{24873952}{6561} \zeta_3
+ \frac{5120}{3} \zeta_4 - \frac{56969}{243} \zeta_5\right] a_s^4
\end{eqnarray}
%----------------------------------------------------------------------------
and can be compared to the Pad\'{e}--approximant
%----------------------------------------------------------------------------
\begin{eqnarray}
\gamma_n^{\rm Pade} &=& \frac{{\gamma_n^{(2)}}^2}{\gamma_n^{(1)}}~.
\end{eqnarray}
%----------------------------------------------------------------------------
For the second moment we found a deviation of about 20\% only in the 
$O(a_s^4)$ term, which is far inside our uncertainty margins in the 
non--singlet case. The variation in $\Lambda_{\rm QCD}$ varying the 4--loop 
anomalous dimension between zero and twice the value of the Pad\'{e}--approximant 
amounts to $\pm 2 \MeV$. Thus $\Lambda_{\rm QCD}$ at $O(a_s^4)$ accuracy is 
widely determined by the 3--loop Wilson coefficient and the behaviour of $a_s(Q^2)$ 
to 4--loop order.

Table~5 summarizes our fit results comparing  $\Lambda_{\rm QCD}^{\rm N_f =4}$ 
and $\alpha_s(M_Z^2)$ for  the NLO, NNLO and N$^3$LO analysis.
The NNLO value of $\Lambda_{\rm QCD}^{\rm N_f =4}$ is found to be smaller 
than the NLO value, while the N$^3$LO value comes out somewhat higher
than the NNLO value at about the same experimental error. The pattern
of central values from NLO to N$^3$LO shows the convergence of the 
analysis, which is illustrated by the decreasing difference of these values. 
This measure of convergence is more significant than 
varying the scales of the problem, which is less well defined since the 
range of variations is arbitrary.  

We compare the results of the present analysis to results obtained in 
the literature at  NLO and NNLO in Table~6. Most of the NLO values for 
$\alpha_s(M_Z^2)$ shown were determined in combined singlet- and 
non--singlet analyses, partly including jet--data in $\overline{p} p$ 
scattering. Not in all analyses the correlation matrices for the parameters of
the fit were presented. The NLO values for $\alpha_s(M_Z^2)$ are larger than 
those at NNLO as several analyses, which performed both fits, show. The 
difference of both values, however, is not always the same, most likely 
due to the type of analysis being performed (singlet and non-singlet, 
non-singlet only, etc.), in which also partly different data sets are analyzed.

As outlined above  we perform kinematic cuts in the present analysis to 
eliminate as widely as possible higher twist effects. In this way a
significant part of the SLAC data is cut away due to the requirement of a 
minimal value $W^2 = 12.5 \GeV^2$. Other analyses, cf. 
\cite{A02,Gluck:2006yz}, 
fit {\sf phenomenological models} for the higher twist contributions. In a 
precision determination of $\alpha_s(M_Z^2)$ we do not follow this way since a 
thorough description of these contributions in terms of the light-cone 
expansion, including the description of the 4--, 6-- etc. parton 
correlation functions, the corresponding Wilson coefficients and anomalous 
dimensions, is not known yet, but would be required. 

A significant part of the data in the non--singlet case is due to the 
BCDMS measurements
\cite{BCDMS}. These data have repeatedly being criticized to deliver a too 
small value for $\Lambda_{\rm QCD}$, respectively $\alpha_s(M_Z^2)$. Let us 
therefore revisit associated analyses. The NLO BCDMS--fit was confirmed
in \cite{KRI2} both in the singlet and a separate non--singlet analysis 
with practically the same value for 
$\Lambda_{\rm QCD}^{\rm N_f =4,\overline{\rm MS}}$,
%------------------------------------------------------------------------
\begin{eqnarray}
\Lambda_{\rm QCD}^{\rm N_f =4,\overline{\rm MS}} = 198 \pm 20 \MeV~;~~
\alpha_s(M_Z^2) = 0.1096 \footnotesize{\begin{array}{c} +0.0016 \\ 
-0.0018 \end{array}}~.
\end{eqnarray}
%------------------------------------------------------------------------
In a re-analysis of the BCDMS data \cite{KRI3} using somewhat less severe 
$y$ cuts than applied in the present paper the corresponding value  
%------------------------------------------------------------------------
\begin{eqnarray}
\Lambda_{\rm QCD}^{\rm N_f =4,\overline{\rm MS}} = 257 \pm 40 \MeV,
\end{eqnarray}
%------------------------------------------------------------------------
was found, which is numerically in accordance with our result. However, in
\cite{KRI3} the flavor threshold matching in $\alpha_s(\mu^2)$ was 
performed at the rather high scales $4 m_Q^2,~~m_c^2 = 9 \GeV^2, m_b^2 = 
80 \GeV^2$. Usually the convention is to match at $m_c$ and $m_b$. 
Therefore, the results of \cite{KRI3} cannot be compared to other 
analyses directly. We preformed the NLO analysis with these
matching conditions and obtained
%------------------------------------------------------------------------
\begin{eqnarray}
\Lambda_{\rm QCD}^{\rm N_f =4,\overline{\rm MS}} = 254 \pm 26 \MeV;
~~\alpha_s(M_Z^2) = 0.1127 \footnotesize{\begin{array}{c} +0.0018 \\
-0.0019 \end{array}}
\end{eqnarray}
%------------------------------------------------------------------------
and do not confirm the rather high values of $\alpha_s(M_Z^2)$ reported in 
\cite{KRI3}.

Non--singlet QCD analyses were also performed for neutrino data. In 
\cite{KAT} the CCFR iron data on $xF_3(x,Q^2)$ \cite{Seligman:1997mc} 
were analyzed in NLO and 
NNLO using fixed moments. Likewise a NNLO analysis was performed in 
\cite{SY}. Iron data suffer form the EMC effect, Fermi motion and other 
nuclear effects, as re-scattering,  for which the QCD scaling violations  
are not known in detail. In \cite{KAT} rather large values for
$\Lambda_{\rm QCD}^{\rm N_f =4,\overline{\rm MS}}$ 
%------------------------------------------------------------------------
\begin{eqnarray}
\Lambda_{\rm QCD,NLO}^{\rm N_f =4,\overline{\rm MS}} = 371 \pm 72 \MeV\\
\Lambda_{\rm QCD,NNLO}^{\rm N_f =4,\overline{\rm MS}} = 316 \pm 51 \MeV
\end{eqnarray}
%------------------------------------------------------------------------
are obtained, which are  higher than the values obtained in the 
analyses based on $F_2^{p,d}(x,Q^2)$ data, still showing the pattern that
the NNLO value is lower than the NLO value.
In a quite similar analysis in Ref. \cite{SY} on the other hand, values close 
to the results of the present analysis are obtained,
%------------------------------------------------------------------------
\begin{eqnarray}
\Lambda_{\rm QCD,NLO}^{\rm N_f =4,\overline{\rm MS}}  &=& 281 \pm 57 \MeV\\
\Lambda_{\rm QCD,NNLO}^{\rm N_f =4,\overline{\rm MS}} &=& 255 \pm 55 \MeV~.
\end{eqnarray}
%------------------------------------------------------------------------

In Figure~13 we compare the results of different NLO and NNLO QCD 
unpolarized and polarized DIS analyses for $\alpha_s(M_Z^2)$ with the 
results obtained in the present paper and the world average for
$\alpha_s(M_Z^2)$ 
%------------------------------------------------------------------------
\begin{eqnarray}
\alpha_s(M_Z^2) = 0.1189 \pm 0.0010~, 
\end{eqnarray}
%------------------------------------------------------------------------
which has been determined in \cite{Bethke:2006ac} recently.
In most of the cases the central values lay below the world average.
Yet most of the NLO results are 1$\sigma$ compatible with the world 
average and an increasing number of NNLO results departs to lower
values. In the case of polarized deep inelastic scattering at present
only NLO analyses are performed with larger errors than in the 
unpolarized case, which is due the current experimental errors of the 
polarization asymmetries.  

We finally would like to mention results on 
$\Lambda_{\rm QCD}^{\rm N_f =2,\overline{\rm MS}}$ from lattice 
calculations with two dynamical quark flavors 
\cite{DellaMorte:2004bc,Gockeler:2005rv}
%------------------------------------------------------------------------
\begin{eqnarray}
\Lambda_{\rm QCD}^{\rm N_f =2,\overline{\rm MS}} &=& 245 \pm 16 {\rm 
(stat.)} \pm 16 {\rm (syst.)} \MeV\\
\Lambda_{\rm QCD}^{\rm N_f =2,\overline{\rm MS}} &=& 261 \pm 17 {\rm 
(stat.)} \pm 26 {\rm (syst.)} \MeV~.
\end{eqnarray}
%------------------------------------------------------------------------
These values are very close to the $N_f = 0$ lattice results with rather small 
errors. It will be interesting to see which values will be obtained for 
four dynamical quark flavors in the future.
%%%%%%%%%%%%%%%%%%%%%%%%%%%%%%%%%%%%%%%%%%%%%%%%%%%%%%%%%%%%%%%%%%%%%%%
%	higher twist
%%%%%%%%%%%%%%%%%%%%%%%%%%%%%%%%%%%%%%%%%%%%%%%%%%%%%%%%%%%%%%%%%%%%%%%
\section{Higher Twist}
\label{sec:ht}
%%%%%%%%%%%%%%%%%%%%%%%%%%%%%%%%%%%%%%%%%%%%%%%%%%%%%%%%%%%%%%%%%%%%%%%

\vspace{1mm}\noindent
The cuts applied in $Q^2$ and $W^2$, $Q^2 \geq 4 \GeV^2$ and $W^2 \geq 
12.5 \GeV^2$ in the standard analysis do widely eliminate higher twist 
effects from the data sample. Still the region for deep--inelastic 
scattering ranges to lower values of $W^2 \geq 4 \GeV^2$ keeping the cut
in $Q^2$. The data in the kinematic region 
%-----------------------------------------------------------------------
\begin{eqnarray}
\label{eqHT}
Q^2 \geq 4 \GeV^2,~~~~~~4 \leq W^2 \leq 12.5 \GeV^2 
\end{eqnarray}
%-----------------------------------------------------------------------
may thus be used to yield information on power corrections in 
$F_2^p(x,Q^2)$ and $F_2^d(x,Q^2)$. For this purpose we extrapolate the QCD 
fit results obtained for $W^2 \geq 12.5 \GeV^2$ to the region (\ref{eqHT})
and form the difference between data and theory, applying target mass 
corrections in addition. The empirical higher twist coefficient
%-----------------------------------------------------------------------
\begin{eqnarray}
\label{eqHTC}
F_2^{\rm exp}(x,Q^2) = O_{\rm TMC}[F_2^{\rm tw2}(x,Q^2)] \cdot
\left( 1 + \frac{C_{\rm HT}(x,Q^2)}{Q^2[\GeV^2]}\right)
\end{eqnarray}
%-----------------------------------------------------------------------
is extracted. Here the operation $O_{\rm TMC}[...]$ denotes taking the 
target mass corrections of the twist--2 contributions to the 
respective structure function. The coefficients $C_{\rm HT}(x,Q^2)$ are
determined in bins of $x$ and $Q^2$ and are then averaged over $Q^2$. The extracted
distributions for $C_{\rm HT}(x)$ are depicted in Figures~14 and 15
for the non--singlet case considering scattering off the proton and 
the deuteron target, respectively. The coefficient $C_{\rm HT}(x)$ grows 
towards large $x$. We find a similar pattern if compared to the early 
NLO analysis \cite{VM92}. Comparing the extractions for the case of 
calculating $F_2^{\rm tw2}(x,Q^2)$ in NLO, NNLO and N$^3$LO a gradual 
reduction of $C_{\rm HT}(x)$ is obtained, i.e. higher twist extractions 
describing the twist--2 contributions by low order in perturbation theory 
only lead to incorrect results. As seen from Figures~14, 15 $C_{\rm HT}(x)$ 
is widely independent of the target comparing the results for deeply 
inelastic scattering off protons and deuterons, after nuclear wave function 
corrections were performed. Beyond $x \simeq 0.9$ the errors for 
$C_{\rm HT}(x)$ become very large.
%%%%%%%%%%%%%%%%%%%%%%%%%%%%%%%%%%%%%%%%%%%%%%%%%%%%%%%%%%%%%%%%%%%%%%%
%	conclusions
%%%%%%%%%%%%%%%%%%%%%%%%%%%%%%%%%%%%%%%%%%%%%%%%%%%%%%%%%%%%%%%%%%%%%%%
\section{Conclusions}
\label{sec:concl}
%%%%%%%%%%%%%%%%%%%%%%%%%%%%%%%%%%%%%%%%%%%%%%%%%%%%%%%%%%%%%%%%%%%%%%%

\vspace{1mm}\noindent
We performed a  twist--2 QCD analysis of the non--singlet world data up to
NNLO and N$^3$LO and determined the valence quark densities $xu_v(x,Q^2)$
and $xd_v(x,Q^2)$ with correlated errors in the $\overline{\rm MS}$--scheme
using conservative kinematic cuts for the data.
Parameterizations of these parton distribution functions and their errors 
were derived in a wide range of $x$ and $Q^2$  as fit results at 
LO, NLO, NNLO, and N$^3$LO. In the analysis the QCD scale $\Lambda_{\rm QCD}^{\rm N_f = 4}$ 
and the strong coupling constant $\alpha_s(M_Z^2)$, were determined up
to N$^3$LO. The effect of the 4--loop anomalous dimension, if compared 
to the 3--loop Wilson coefficient and the 4--loop effects in $\alpha_s(\mu^2)$   
turn out to be rather weak. The corresponding estimates, based on a Pad\'{e}-approximation
compare very well with the recently calculated second moment of the 4--loop
anomalous dimension. The assumption of a $\pm 100\%$ error on the Pad\'{e}-approximant
results into a marginal shift of $\pm 2 \MeV$ in $\Lambda_{\rm QCD}^{\rm N_f = 4}$
only. The convergence of the central values for $\Lambda_{\rm QCD}^{\rm N_f = 4}$
and $\alpha_s(M_Z^2)$ from NLO to N$^3$LO shows the stability of the analysis
and provides some measure of remaining higher order effects.
Moments for the valence quark distributions with correlated errors at NNLO
and N$^3$LO were calculated for comparisons with upcoming lattice simulations.
Higher twist effects both in $F_2^p$ and $F_2^d$ are extracted extrapolating
the twist--2 results to the region $W^2 \leq 12.5 \GeV^2$. These contributions
are found to be widely target independent and decrease with rising order
in the coupling constant for the twist--2 contributions.  

%%%%%%%%%%%%%%%%%%%%%%%%%%%%%%%%%%%%%%%%%%%%%%%%%%%%%%%%%%%%%%%%%%%%%%%
% Acknowledgements
%%%%%%%%%%%%%%%%%%%%%%%%%%%%%%%%%%%%%%%%%%%%%%%%%%%%%%%%%%%%%%%%%%%%%%%
%\newpage

\vspace{3mm}\noindent
{\bf Acknowledgment}.~This work was supported in part by DFG
Sonderforschungsbereich Transregio 9, Computergest\"utzte Theoretische Physik. 
We thank J. Stirling for the kind permission to show Figure~1 and for discussions.
We would like to thank E.~Reya and S. Alekhin  for communicating their recent 
results  to us. We would like to thank E.~Christy, R.~Ent, M.~Gl\"uck, 
K.~Jansen, C.~Keppel, M.~Klein, W.~Melnitchouk, S.~Moch, G.~Schierholz, R.~Sommer, 
A.W.~Thomas and U.~Wolff for discussions and DESY for support of this project. 

\vspace{2mm}
\noindent
\newpage

%%%%%%%%%%%%%%%%%%%%%%%%%%%%%%%%%%%%%%%%%%%%%%%%%%%%%%%%%%%%%%%%%%%%%%%
%	Tables
%%%%%%%%%%%%%%%%%%%%%%%%%%%%%%%%%%%%%%%%%%%%%%%%%%%%%%%%%%%%%%%%%%%%%%%
\section{Tables}
\label{sec:tab}
%%%%%%%%%%%%%%%%%%%%%%%%%%%%%%%%%%%%%%%%%%%%%%%%%%%%%%%%%%%%%%%%%%%%%%%
%%%%%%%%%%%%%%%%%%%%%%%%%%%%%%%%%%%%%%%%%%%%%%%%%%%%%%%%%%%%%%%%%%%%%%%%
%%%%%%%%%%%%%%%%%%%%%%%%%%%%%%%%%%%%%%%%%%%%%%%%%%%%%%%%%%%%%%%%%%%%%%%%
%
\renewcommand{\arraystretch}{1.3}
\normalsize
\begin{center}
\begin{tabular}{|l|c|c|r|r|r|r|}
\hline \hline
Experiment  & $x$ & $Q^2,~\GeV^2$ & $F_2^p$ & $F_2^p~cuts$ & $F_2^p~HT$ & Norm  \\
\hline \hline
BCDMS (100)   & 0.35 -- 0.75 &  11.75 --  75.00  &  51 &  21 &  10 &
1.005 \\  
BCDMS (120)   & 0.35 -- 0.75 &  13.25 --  75.00  &  59 &  32 &   4 &
0.998 \\
BCDMS (200)   & 0.35 -- 0.75 &  32.50 -- 137.50  &  50 &  28 &   0 &
0.998 \\
BCDMS (280)   & 0.35 -- 0.75 &  43.00 -- 230.00  &  49 &  26 &   0 &
0.998 \\
NMC (comb)    & 0.35 -- 0.50 &   7.00 --  65.00  &  15 &  14 &   6 &
1.000 \\
SLAC (comb)   & 0.30 -- 0.62 &   7.30 --  21.39  &  57 &  57 & 259 &
1.013 \\
H1 (hQ2)      & 0.40 -- 0.65 &    200 --  30000  &  26 &  26 &   0 &
1.020 \\
ZEUS (hQ2)    & 0.40 -- 0.65 &    650 --  30000  &  15 &  15 &   0 &
1.007 \\
\hline
$proton$      &              &                   & 322 & 227 & 279 &  \\
\hline \hline
Experiment  & $x$ & $Q^2, \GeV^2$ & $F_2^d$ & $F_2^d~cuts$ & $F_2^d~~HT$ & Norm  \\
\hline \hline
BCDMS (120)   & 0.35 -- 0.75 & 13.25 --  99.00  &  59 &  32 &   4 & 1.001 \\
BCDMS (200)   & 0.35 -- 0.75 & 32.50 -- 137.50  &  50 &  28 &   0 & 0.998 \\
BCDMS (280)   & 0.35 -- 0.75 & 43.00 -- 230.00  &  49 &  26 &   0 & 1.003 \\
NMC (comb)    & 0.35 -- 0.50 &  7.00 --  65.00  &  15 &  14 &   6 & 1.000 \\
SLAC (comb)   & 0.30 -- 0.62 & 10.00 --  21.40  &  59 &  59 & 268 & 0.990 \\
\hline
$deuteron$    &              &                  & 232 & 159 & 278 & \\
\hline \hline
Experiment  & $x$ & $Q^2, \GeV^2$ & $F_2^{NS}$ & $F_2^{NS}~(cuts)$ & $F_2^{NS}~~HT$
& Norm  \\
\hline \hline
BCDMS (120)  & 0.070 -- 0.275 &  8.75 --  43.00  &  36 &  30 &   0 &  0.983 \\
BCDMS (200)  & 0.070 -- 0.275 & 17.00 --  75.00  &  29 &  28 &   0 &  0.999 \\
BCDMS (280)  & 0.100 -- 0.275 & 32.50 -- 115.50  &  27 &  26 &   0 &  0.997 \\
NMC (comb)   & 0.013 -- 0.275 &  4.50 --  65.00  &  88 &  53 &   0 &  1.000 \\
SLAC (comb)  & 0.153 -- 0.293 &  4.18 --   5.50  &  28 &  28 &   1 &  0.994 \\
\hline
$non-singlet$ &               &                  & 208 & 165 &   1 & \\
\hline \hline
$total$       &               &                  & 762 & 551 & 558 & \\  
\hline \hline
\end{tabular}
\end{center}
\normalsize
\vspace{2mm}
\noindent
%\caption{\label{table:data}Number of data points for the non--singlet
%QCD analysis 
{\sf Table~1: Number of data points for the non--singlet QCD analysis
with their $x$ and $Q^2$ ranges. In the first column are given (in
parentheses) the beam momentum in GeV of the the respective data set (number), a flag 
whether the data come from a combined analysis of all beam momenta (comb) 
or whether the data are taken at high momentum transfer (hQ2). The fourth column
($F_2$) contains the number of data points according to the cuts: $Q^2
> 4~{\GeV^2}$, $W^2 > 12.5~{ \GeV^2}$, $x > 0.3$ for $F_2^p$ and
$F_2^d$ and $x < 0.3$ for $F_2^{NS}$. The reduction of the number of
data points by the additional cuts on the BCDMS data ($y > 0.3$) and
on the NMC data ($Q^2 > 8~{ \GeV^2}$) are given in the 5th column
($F_2~cuts$). The 6th column ($F_2~HT$) contains the number of data
points in the range $4~{\GeV^2} < W^2 < 12.5~{\GeV^2}$ used to
fit the higher twist coefficients $C_{HT}(x)$ for the proton and
deuteron data. In the last column the normalization shifts (see text)
are listed. 
\renewcommand{\arraystretch}{1.0}
%
%%%%%%%%%%%%%%%%%%%%%%%%%%%%%%%%%%%%%%%%%%%%%%%%%%%%%%%%%%%%%%%%%%%%%%%%
%%%%%%%%%%%%%%%%%%%%%%%%%%%%%%%%%%%%%%%%%%%%%%%%%%%%%%%%%%%%%%%%%%%%%%%%
%
\renewcommand{\arraystretch}{1.3}
\normalsize
\begin{center}
\begin{tabular}{|c|c|c|c|c|}
\hline \hline 
           &             & NLO  & NNLO & N$^3$LO \\
\hline
$u_v$      & $a$         &  0.274 $\pm$ 0.027 &  0.291 $\pm$ 0.008 & 
 0.298 $\pm$ 0.008 \\
           & $b$         &  3.909 $\pm$ 0.040 &  4.013 $\pm$ 0.037 & 
 4.032 $\pm$ 0.037 \\
           & $\rho  $    &  6.003             &  6.227             & 
 6.042             \\
           & $\gamma$    & 35.089             & 35.629             & 
35.492             \\
\hline
$d_v$      & $a$         &  0.461 $\pm$ 0.030 &  0.488 $\pm$ 0.033 & 
 0.500 $\pm$ 0.034 \\
           & $b$         &  5.683 $\pm$ 0.228 &  5.878 $\pm$ 0.239 & 
 5.921 $\pm$ 0.243 \\
           & $\rho  $    & --3.699             & --3.639             & 
--3.618             \\
           & $\gamma$    & 16.491             & 16.445             & 
16.414             \\
\hline
\multicolumn{2}{|c|}{$\Lambda_{\rm QCD}^{\rm N_f=4}$, MeV} & 265 $\pm$ 27 & 
226 $\pm$ 25 & 234 $\pm$ 26 \\
\hline \hline
\multicolumn{2}{|c|}{$\chi^2 / ndf$} & 484/546 = 0.89 & 472/546 = 0.86 & 
461/546 = 0.84 \\
\hline \hline            
\end{tabular}
\end{center}
\normalsize
\vspace{2mm}
\noindent
{\sf Table~2: Parameter values of the NLO, NNLO and N$^3$LO non--singlet
QCD fit at $Q_0^2 = 4~\GeV^2$. The values without error have been fixed
after a first minimization since the data do not constrain these
parameters well enough (see text).}   
\renewcommand{\arraystretch}{1.0}

\vspace{4cm}
%
%%%%%%%%%%%%%%%%%%%%%%%%%%%%%%%%%%%%%%%%%%%%%%%%%%%%%%%%%%%%%%%%%%%%%%%%
%
\renewcommand{\arraystretch}{1.2}
\normalsize
\begin{center}
\begin{tabular}{|c||c|c|c|c|c|}
\hline \hline
 NNLO & $\Lambda_{\rm QCD}^{\rm N_f=4}$ & $a_{u_v}$ & $b_{u_v}$ & $a_{d_v}$ & $b_{d_v}$ \\
\hline \hline
 $\Lambda_{QCD}^{(4)}$ & {\bf 6.45E-4} &  &  &  &  \\
\hline
 $a_{u_v}$        &  9.03E-5 & {\bf 5.75E-5} &  &  &  \\
\hline
 $b_{u_v}$        & -3.37E-4 &  1.55E-4 & {\bf 1.40E-3} &  & \\
\hline
 $a_{d_v}$        &  1.92E-4 & -8.97E-6 & -4.69E-4 & {\bf 1.07E-3} & \\ 
\hline
 $b_{d_v}$        &  9.19E-4 &  5.82E-5 & -3.30E-3 & 7.21E-3 & {\bf 5.72E-2} \\ 
\hline
\hline
\end{tabular} 
\end{center}
\normalsize
\vspace{2mm}
\noindent
\begin{center}
{\sf Table~3: The covariance matrix of the NNLO non--singlet QCD fit at
$Q_0^2 = 4~\GeV^2$.}   
\end{center}
\renewcommand{\arraystretch}{1.0}
\vspace{5mm}

\newpage
%
%%%%%%%%%%%%%%%%%%%%%%%%%%%%%%%%%%%%%%%%%%%%%%%%%%%%%%%%%%%%%%%%%%%%%%%%
%

\vspace*{6cm}
\renewcommand{\arraystretch}{1.3}
\normalsize
\begin{center}
\begin{tabular}{|c|c|c|c|c|c|c|}
\hline \hline 
 $f$ & $n$ & N$^3$LO & NNLO & MRST04 & A02 & A06 \\
\hline \hline
$u_v$ & 2 & 0.3006 $\pm$ 0.0031 & 0.2986 $\pm$ 0.0029 & 0.285 & 0.304 & 0.2947\\
%------
      & 3 & 0.0877 $\pm$ 0.0012 & 0.0871 $\pm$ 0.0011 & 0.082 & 0.087 & 0.0843 \\ 
%------
      & 4 & 0.0335 $\pm$ 0.0006 & 0.0333 $\pm$ 0.0005 & 0.032 & 0.033 & 0.0319 \\
%------
\hline
%----------------------------------------------------------------------
$d_v$ & 2 & 0.1252 $\pm$ 0.0027 & 0.1239 $\pm$ 0.0026 & 0.115 & 0.120 & 0.1129\\
%------
      & 3 & 0.0318 $\pm$ 0.0009 & 0.0315 $\pm$ 0.0008 & 0.028 & 0.028 & 0.0275\\ 
%------
      & 4 & 0.0106 $\pm$ 0.0004 & 0.0105 $\pm$ 0.0004 & 0.009 & 0.010 & 0.0092\\
%------
\hline
%----------------------------------------------------------------------
$u_v-d_v$ & 2 & 0.1754 $\pm$ 0.0041 & 0.1747 $\pm$ 0.0039 & 0.171 & 0.184 & 0.182\\ 
%------
          & 3 & 0.0559 $\pm$ 0.0015 & 0.0556 $\pm$ 0.0014 & 0.055 & 0.059 & 0.057\\ 
%------
          & 4 & 0.0229 $\pm$ 0.0007 & 0.0228 $\pm$ 0.0007 & 0.022 & 0.024 & 0.023\\
%------
\hline \hline
\end{tabular}
\end{center}
\normalsize
\vspace{2mm}
\noindent
{\sf Table~4: Comparison of low order moments at $Q_0^2 =
4~\GeV^2$ from our non--singlet N$^3$LO and NNLO QCD analyses with
the NNLO analyses MRST04~\cite{MRST04}, A02~\cite{A02} and A06~\cite{Alekhin:2006zm}
derived from global analyses, resp. combined singlet/non-singlet fits.}      
\renewcommand{\arraystretch}{1}

\newpage
%
%%%%%%%%%%%%%%%%%%%%%%%%%%%%%%%%%%%%%%%%%%%%%%%%%%%%%%%%%%%%%%%%%%%%%%%
%
\renewcommand{\arraystretch}{1.3}
\normalsize
\begin{center}
\begin{tabular}{|c|c||c|}
\hline
\hline 
   & $\Lambda_{\rm QCD}^{\rm N_f=4}$, MeV & $\alpha_s(M_Z^2)$ \\
\hline
 NLO  &  265 $\pm$ 27 & 0.1148 $\begin{array}{c} +0.0019 \\ -0.0019
\end{array}$ (expt) \\ 
 NNLO &  226 $\pm$ 25 & 0.1134 $\begin{array}{c} +0.0019 \\ -0.0021
\end{array}$ (expt) \\ 
 N$^3$LO &  234 $\pm$ 26 & 0.1141 $\begin{array}{c} +0.0020 \\ -0.0022
\end{array}$ (expt) \\ 
\hline
\hline
\end{tabular}
\end{center}
\normalsize
\vspace{2mm}
\noindent
\begin{center}
{\sf Table~5: $\Lambda_{\rm QCD}^{\rm N_f=4}$ and $\alpha_s(M_Z^2)$ at NLO, NNLO
and N$^3$LO.}   
\end{center}
\renewcommand{\arraystretch}{1.00}
%
%%%%%%%%%%%%%%%%%%%%%%%%%%%%%%%%%%%%%%%%%%%%%%%%%%%%%%%%%%%%%%%%%%%%%%%%
%
\renewcommand{\arraystretch}{1.1}
\normalsize
\begin{center}
\begin{tabular}{|l|llll|c|}
\hline \hline
                  & $\alpha_s(M_Z^2)$ & $\quad$expt & $\quad$theory &
$\quad$model & Ref. \\ 
\hline
\hline
{\bf  NLO } & & & & & \\
\hline
 CTEQ6  & 0.1165 & $\pm$0.0065 &             &             & \cite{CTEQ_P} \\
 MRST03 & 0.1165 & $\pm$0.0020 & $\pm$0.0030 &             & \cite{MRST03} 
\\
 A02    & 0.1171 & $\pm$0.0015 & $\pm$0.0033 &             & \cite{A02} \\
 ZEUS   & 0.1166 & $\pm$0.0049 &             & $\pm$0.0018 & \cite{ZEUS_Ch} \\
 H1     & 0.1150 & $\pm$0.0017 & $\pm$0.0050 & \hspace*{-0.50cm}
\footnotesize{$\begin{array}{c} +0.0009 \\ -0.0005 \end{array}$} &
\cite{H1} \\   
 BCDMS  & 0.110  & $\pm$0.006 &             &             & \cite{BCDMS} \\  
GRS  &  0.112  &  &            &             &  \cite{Gluck:2006yz}             
\\
BBG  &  0.1148  & $\pm$ 0.0019 &            &             &              
\\
\hline
 BB (pol)     & 0.113 & $\pm$0.004 & \hspace*{-0.50cm}
\footnotesize{$\begin{array}{c} +0.009 \\ -0.006 \end{array}$} & & \cite{BB02}
\\   
\hline
\hline
{\bf  NNLO} & & & & & \\
\hline
 MRST03       & 0.1153 & $\pm$0.0020 & $\pm$0.0030 & & \cite{MRST03} \\
 A02          & 0.1143 & $\pm$0.0014 & $\pm$0.0009 & & \cite{A02} \\
 SY01(ep)     & 0.1166 & $\pm$0.0013 &             & & \cite{SY} \\
 SY01($\nu$N) & 0.1153 & $\pm$0.0063 &             & & \cite{SY} \\
GRS  & 0.111 &  &            &             &       \cite{Gluck:2006yz}       
\\
A06  &  0.1128 & $\pm$ 0.0015 &            &             & \cite{Alekhin:2006zm}            
\\
BBG  &  0.1134  & 
\footnotesize{$\begin{array}{c} +0.0019 \\ -0.0021 \end{array}$} 
& & &\\
\hline
\hline
{\bf  N$^3$LO} & & & & & \\
\hline
BBG  &  0.1141  & 
\footnotesize{$\begin{array}{c} +0.0020 \\ -0.0022 \end{array}$}  
&            &             &\\              
\hline \hline
\end{tabular}
\end{center}
\normalsize
\vspace{2mm}
\noindent
\begin{center}
{\sf Table~6: Comparison of $\alpha_s(M_Z^2)$ values from NLO, NNLO, and N$^3$LO
QCD analyses.}   
\end{center}
\renewcommand{\arraystretch}{1.00}

%%%%%%%%%%%%%%%%%%%%%%%%%%%%%%%%%%%%%%%%%%%%%%%%%%%%%%%%%%%%%%%%%%%%%%%
%	Figures
%%%%%%%%%%%%%%%%%%%%%%%%%%%%%%%%%%%%%%%%%%%%%%%%%%%%%%%%%%%%%%%%%%%%%%%
\newpage
\section{Figures}
\label{sec:fig}
%%%%%%%%%%%%%%%%%%%%%%%%%%%%%%%%%%%%%%%%%%%%%%%%%%%%%%%%%%%%%%%%%%%%%%%

\vspace{2cm}
%Figure~1:~
\begin{figure}[htb]
\begin{center}
\includegraphics[angle=0, width=14.0cm]{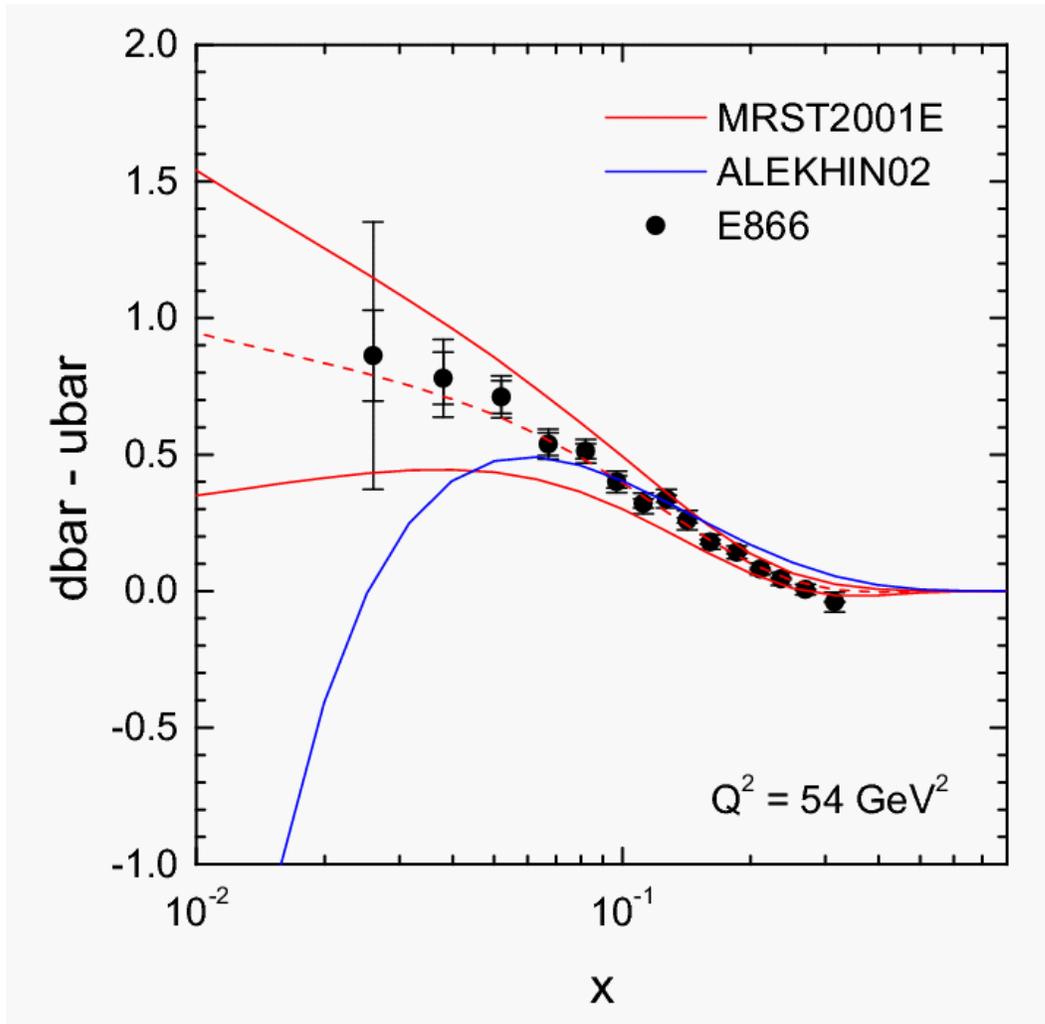} 
\end{center}
\caption{\label{fig:dbmub}
The MRST parameterization of $(\bar{d} - \bar{u})$~\cite{MRST02}
describing the E866 data~\cite{E866}. Here also compared with a
parameterization given in~\cite{A02}.[Courtesy by J. Stirling.]}
\end{figure}
%%%%%%%%%%%%%%%%%%%%%%%%%%%%%%%%%%%%%%%%%%%%%%%%%%%%%%%%%%%%%%%%%%%%%%%%%
%%%%%%%%%%%%%%%%%%%%%%%%%%%%%%%%%%%%%%%%%%%%%%%%%%%%%%%%%%%%%%%%%%%%%%%
\newpage

\vspace*{3cm}
%Figure~2:~
\begin{figure}[htb]
\begin{center}
\includegraphics[angle=0, width=16.0cm]{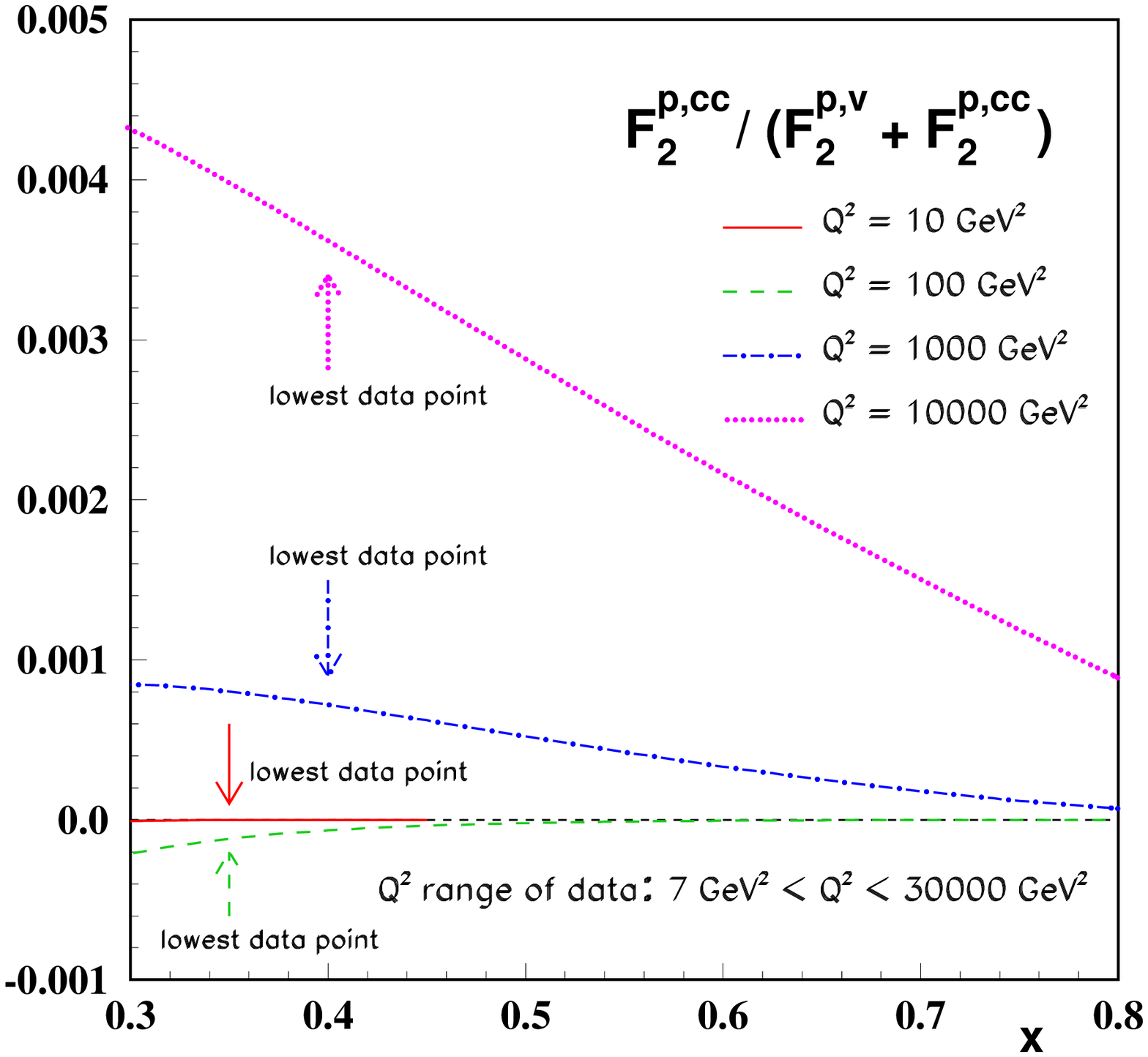} 
\end{center}
\caption{\label{fig:hq1}
Relative size of the NLO contribution due to $c\overline{c}$--production for $m_c = 1.5 \GeV$ for
$F_2^p(x,Q^2)$ in the valence quark region.}
\end{figure}
%%%%%%%%%%%%%%%%%%%%%%%%%%%%%%%%%%%%%%%%%%%%%%%%%%%%%%%%%%%%%%%%%%%%%%%%%
%%%%%%%%%%%%%%%%%%%%%%%%%%%%%%%%%%%%%%%%%%%%%%%%%%%%%%%%%%%%%%%%%%%%%%%
\newpage

\vspace*{3cm}
%Figure~3:~
\begin{figure}[htb]
\begin{center}
\includegraphics[angle=0, width=16.0cm]{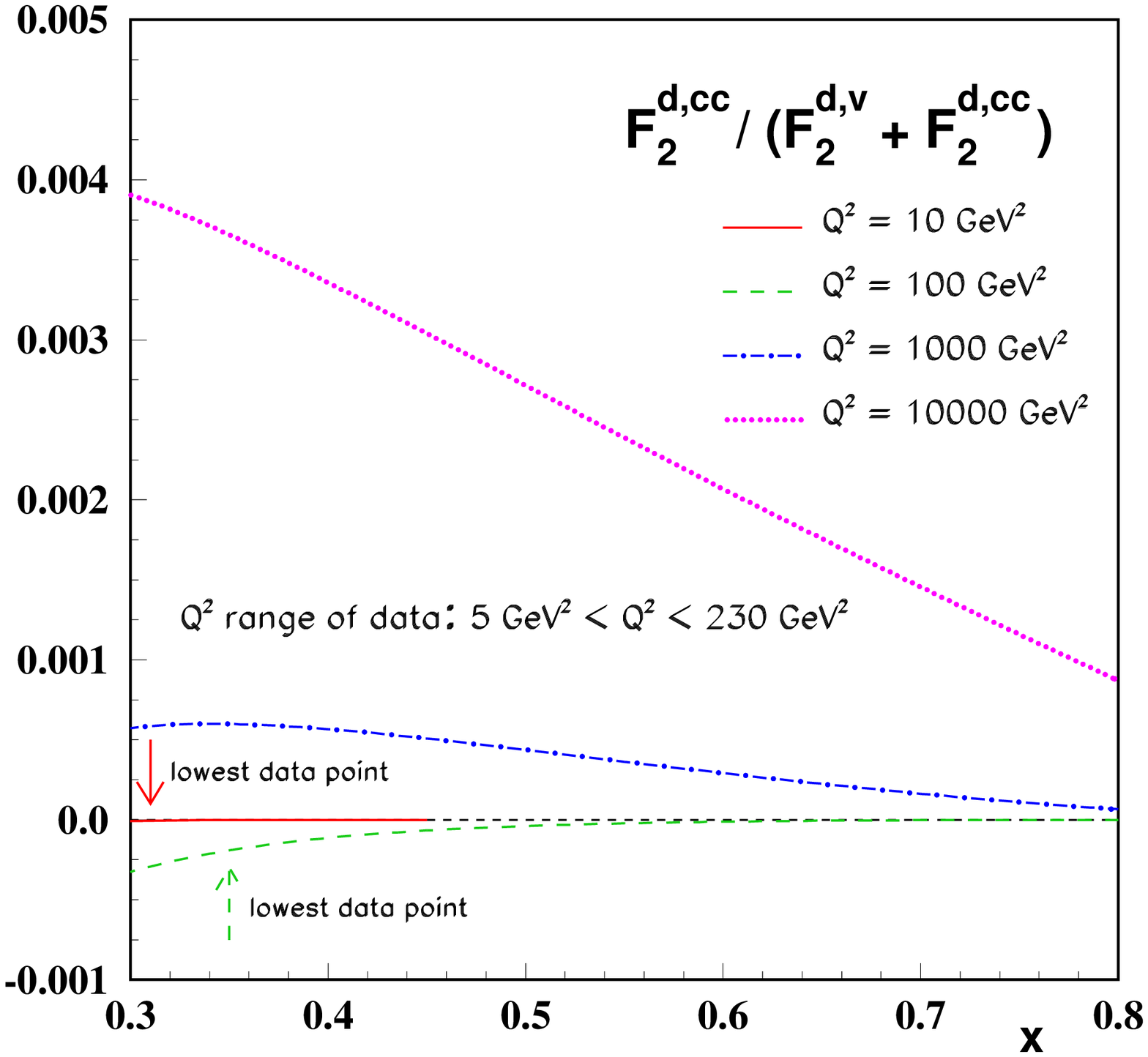} 
\end{center}
\caption{\label{fig:hq2}
Relative size of the NLO contribution due to $c\overline{c}$--production for $m_c = 1.5 \GeV$ for
$F_2^d(x,Q^2)$ in the valence quark region.}
\end{figure}
%%%%%%%%%%%%%%%%%%%%%%%%%%%%%%%%%%%%%%%%%%%%%%%%%%%%%%%%%%%%%%%%%%%%%%%%%
%%%%%%%%%%%%%%%%%%%%%%%%%%%%%%%%%%%%%%%%%%%%%%%%%%%%%%%%%%%%%%%%%%%%%%%
\newpage

\vspace*{2cm}
%Figure~4:~
\vfill
\begin{figure}[htb]
\begin{center}
\includegraphics[angle=0, width=16.0cm]{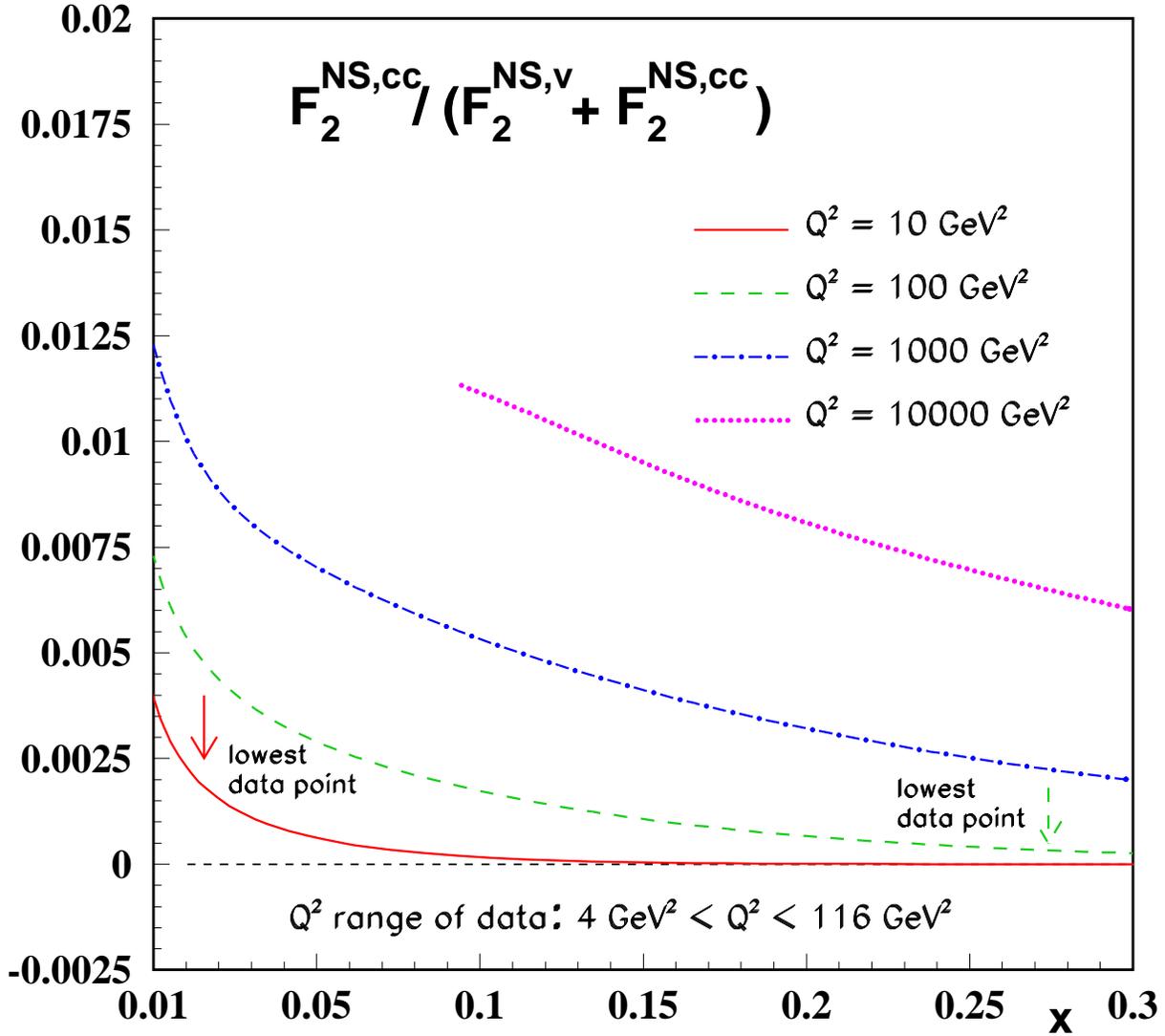} 
\end{center}
\caption{\label{fig:hq3}
Relative size of the NLO contribution due to $c\overline{c}$--production for $m_c = 1.5 \GeV$ for
$F_2^{\rm NS}(x,Q^2)= 2 [F_2^p(x,Q^2) - F_2^d(x,Q^2)]$ in the region $0.01 \leq x \leq 0.3$.}
\end{figure}
\vfill
%%%%%%%%%%%%%%%%%%%%%%%%%%%%%%%%%%%%%%%%%%%%%%%%%%%%%%%%%%%%%%%%%%%%%%%%%
%%%%%%%%%%%%%%%%%%%%%%%%%%%%%%%%%%%%%%%%%%%%%%%%%%%%%%%%%%%%%%%%%%%%%%%
\newpage
%Figure~5:~
%\mbox{\epsfig{file=F2pnsQ2_TMC_HT_nnlo.eps,width=15cm}}
%\vspace{2mm}
%\noindent     
%\small   
\begin{figure}
\begin{center}
\includegraphics[angle=0, width=15.0cm]{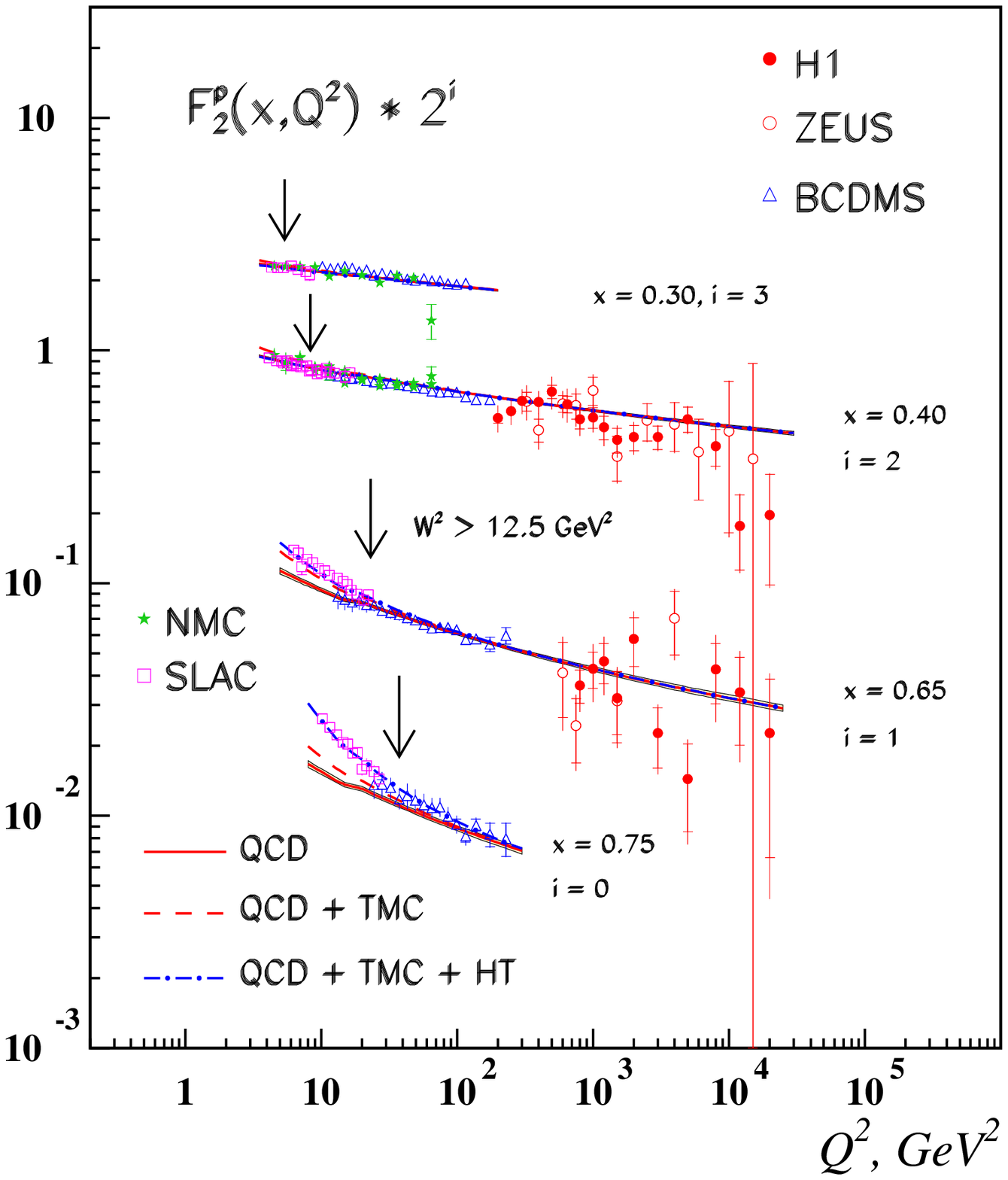} 
\end{center}
{\sf
\caption{\label{fig:f2p_TMC_HT}
The structure function $F_2^p$ as function of $Q^2$ in intervals of
$x$. Shown are the pure QCD fit in NNLO (solid line) and the
contributions from target mass corrections TMC (dashed line) and
higher twist HT (dashed--dotted line). The arrows indicate the
regions with $W^2 > 12.5~{\GeV^2}$. The shaded areas represent the
fully correlated $1\sigma$ statistical error bands.}} 
\end{figure}
%\normalsize
%%%%%%%%%%%%%%%%%%%%%%%%%%%%%%%%%%%%%%%%%%%%%%%%%%%%%%%%%%%%%%%%%%%%%%%
%%%%%%%%%%%%%%%%%%%%%%%%%%%%%%%%%%%%%%%%%%%%%%%%%%%%%%%%%%%%%%%%%%%%%%%
\newpage
%Figure~6:~
%\mbox{\epsfig{file=F2dnsQ2_TMC_HT_nnlo.eps,width=15cm}}
%\vspace{2mm}
%\noindent     
%\small   
\begin{figure}
\begin{center}
\includegraphics[angle=0, width=15.0cm]{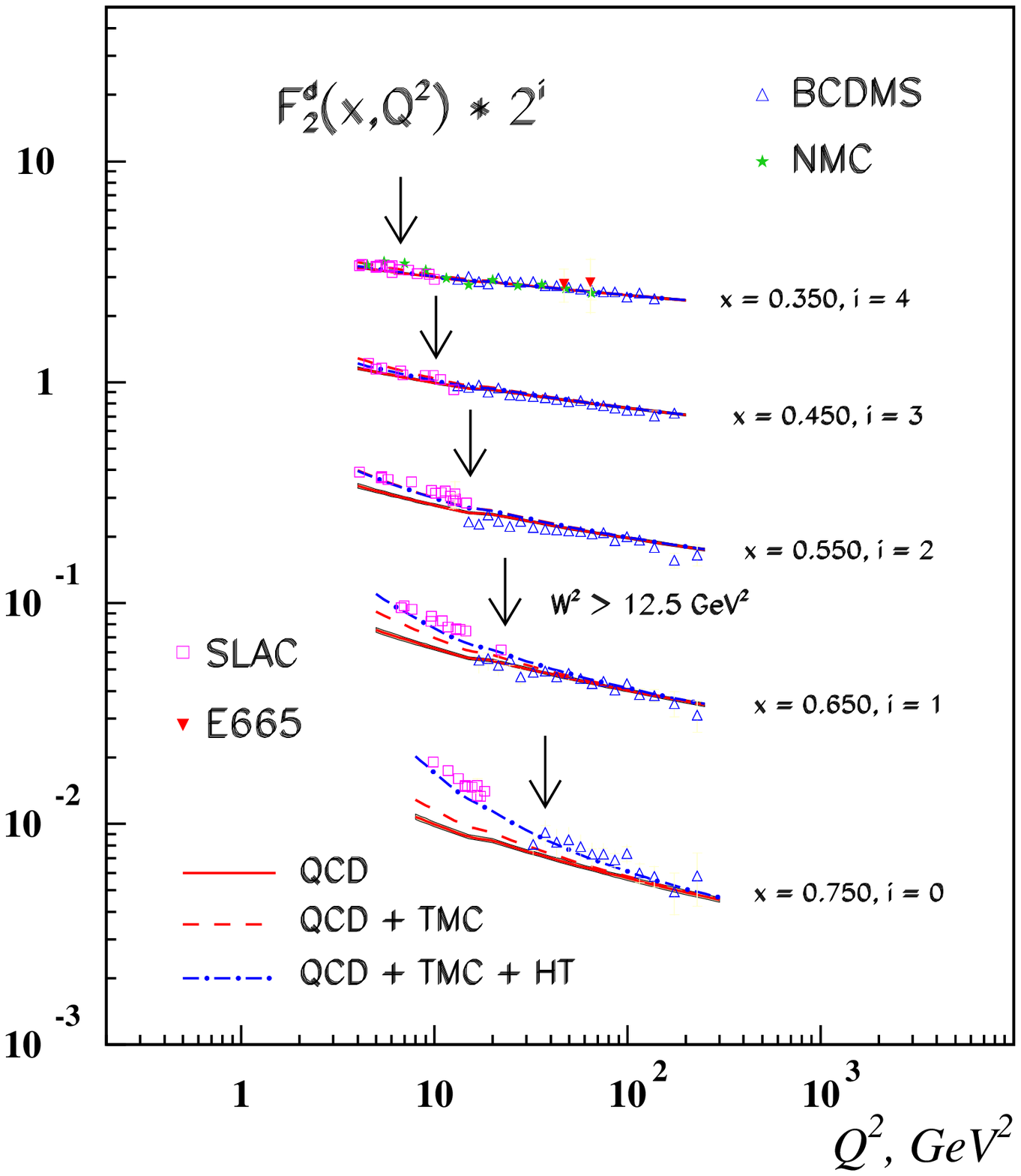} 
\end{center}
{\sf
\caption{\label{fig:f2d_TMC_HT}
The structure function $F_2^d$ as function of $Q^2$ in intervals of
$x$. Shown are the pure QCD fit in NNLO (solid line) and the
contributions from target mass corrections TMC (dashed line) and
higher twist HT (dashed--dotted line). The arrows indicate the
regions with $W^2 > 12.5~{\GeV^2}$. The shaded areas represent the
fully correlated $1\sigma$ statistical error bands.}} 
\end{figure}
%\normalsize
%%%%%%%%%%%%%%%%%%%%%%%%%%%%%%%%%%%%%%%%%%%%%%%%%%%%%%%%%%%%%%%%%%%%%%%
%%%%%%%%%%%%%%%%%%%%%%%%%%%%%%%%%%%%%%%%%%%%%%%%%%%%%%%%%%%%%%%%%%%%%%%
\newpage
%Figure~7:~
%\mbox{\epsfig{file=F2nsvsQ2_nnlo.eps,width=15cm}}
%\vspace{2mm}
%\noindent     
%\small   
\begin{figure}
\begin{center}
\includegraphics[angle=0, width=15.0cm]{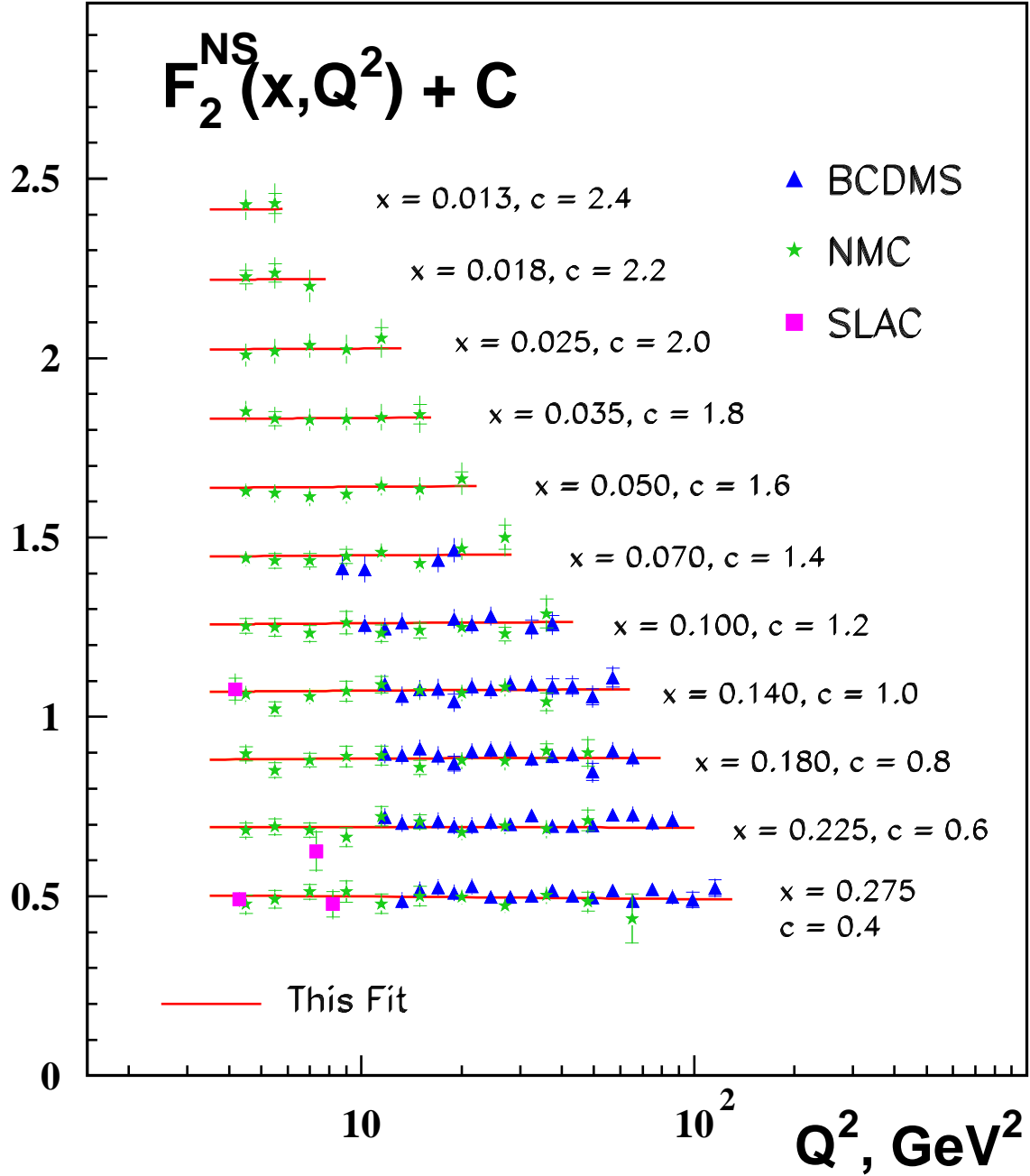} 
\end{center}
{\sf
\caption{\label{fig:f2ns}
The structure function $F_2^{\rm NS}$ as function of $Q^2$ in intervals of
$x$. Shown is the pure QCD fit in NNLO (solid lines) with its fully
correlated $1\sigma$ statistical error band (shaded areas).}}  
\end{figure}
%\normalsize
%%%%%%%%%%%%%%%%%%%%%%%%%%%%%%%%%%%%%%%%%%%%%%%%%%%%%%%%%%%%%%%%%%%%%%%
%%%%%%%%%%%%%%%%%%%%%%%%%%%%%%%%%%%%%%%%%%%%%%%%%%%%%%%%%%%%%%%%%%%%%%%
\newpage
%Figure~8:~
%\mbox{\epsfig{file=F2nstmcQ2_nnlo.eps,width=15cm}}
%\vspace{2mm}
%\noindent     
%\small   
%Figure~7:~
\begin{figure}
\begin{center}
\includegraphics[angle=0, width=15.0cm]{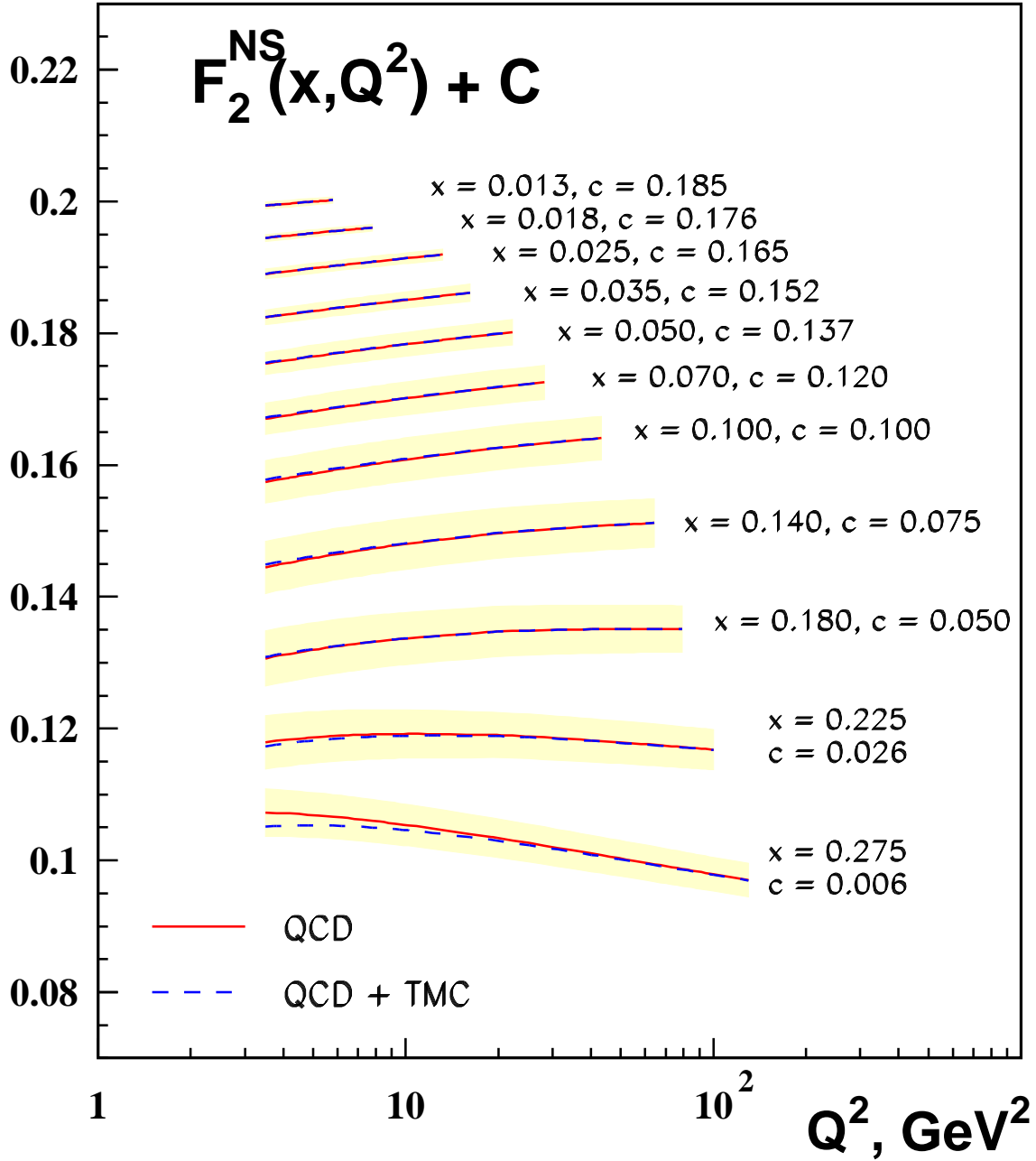} 
\end{center}
{\sf
\caption{\label{fig:f2ns_TMC}
The structure function $F_2^{\rm NS}$ as function of $Q^2$ in intervals of
$x$ without data points. Shown are the pure QCD fit in NNLO (solid
lines) with its fully correlated $1\sigma$ statistical error band
(shaded areas) and the contribution from target mass corrections TMC
(dashed line).}}  
\end{figure}
\normalsize
%%%%%%%%%%%%%%%%%%%%%%%%%%%%%%%%%%%%%%%%%%%%%%%%%%%%%%%%%%%%%%%%%%%%%%%
%%%%%%%%%%%%%%%%%%%%%%%%%%%%%%%%%%%%%%%%%%%%%%%%%%%%%%%%%%%%%%%%%%%%%%%%%
\newpage
%Figure~9:~
%\mbox{\epsfig{file=pdf_comp_nnlo.eps,width=15cm}}
%\vspace{2mm}
%\noindent     
%\small   
%Figure~2:~
\begin{figure}
\begin{center}
\includegraphics[angle=0, width=15.0cm]{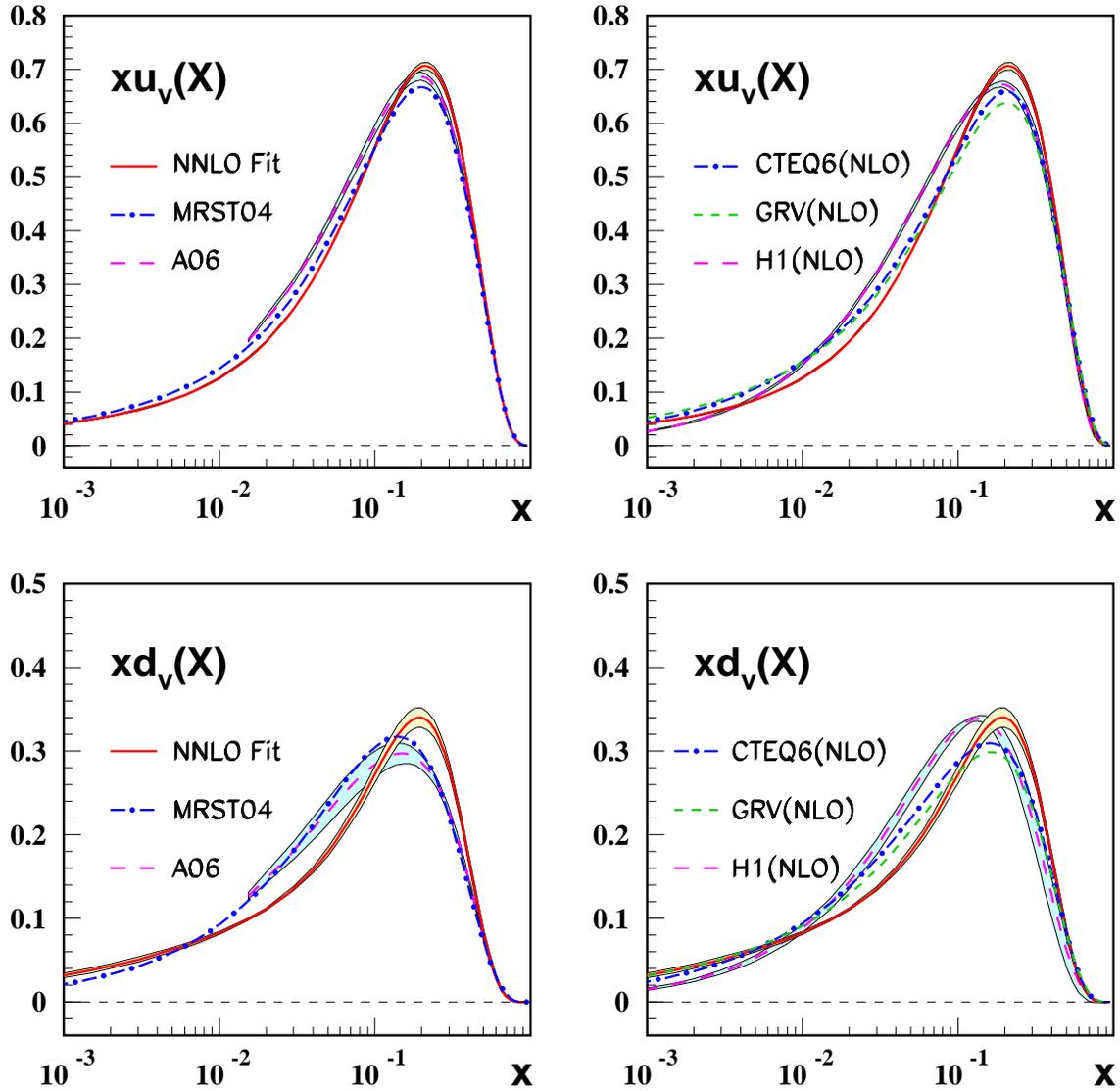} 
\end{center}
{\sf
\caption{\label{fig:pdf_comp1}
The parton densities $xu_v$ and $xd_v$ at the input scale $Q_0^2 =
4.0~{\rm{\GeV^2}}$ (solid line) compared to results obtained from NNLO
analyses by MRST (dashed--dotted line)~\cite{MRST04} and A06
(dashed line)~\cite{Alekhin:2006zm} (left panels) and from NLO analyses by CTEQ6
(dashed--dotted line)~\cite{CTEQ_P}, GRV (dashed line)~\cite{GRV98}
and H1 (long dashed line)~\cite{H1} (right panels). The shaded areas
represent the fully correlated $1\sigma$ statistical error bands.}}
\end{figure}
%\normalsize
%%%%%%%%%%%%%%%%%%%%%%%%%%%%%%%%%%%%%%%%%%%%%%%%%%%%%%%%%%%%%%%%%%%%%%%
%%%%%%%%%%%%%%%%%%%%%%%%%%%%%%%%%%%%%%%%%%%%%%%%%%%%%%%%%%%%%%%%%%%%%%%
\newpage
%Figure~10:~
\begin{figure}
\begin{center}
\includegraphics[angle=0, width=15.0cm]{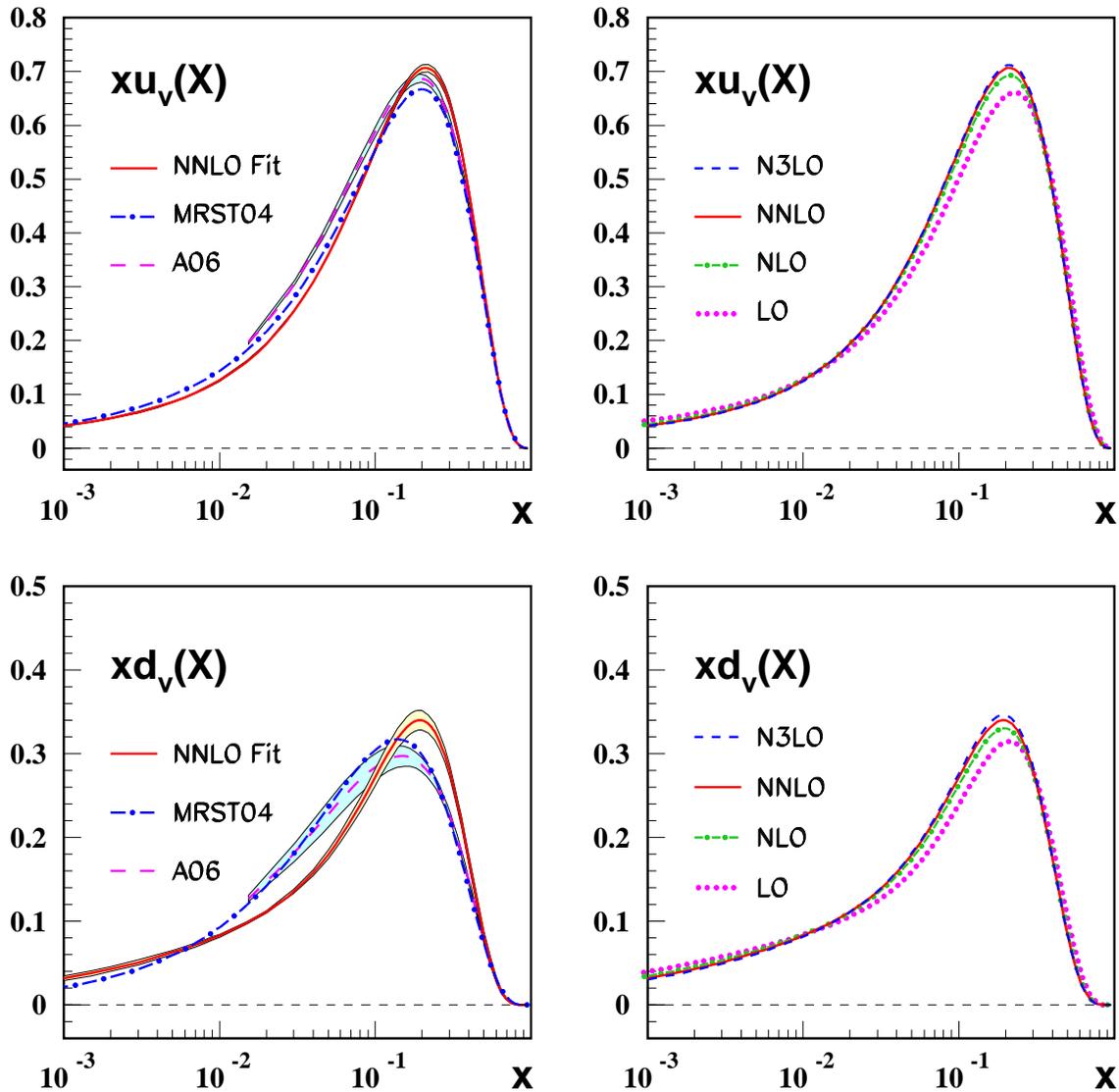} 
\end{center}
{\sf
\caption{%\label{fig:pdf_comp2}
Left panels: The parton densities $xu_v$ and $xd_v$ at the input scale
$Q_0^2 = 4.0~\GeV^2$ (solid line) compared to results obtained
from NNLO analyses by MRST (dashed--dotted line)~\cite{MRST04} and
A06 (dashed line)~\cite{Alekhin:2006zm}. The shaded areas represent the fully
correlated $1\sigma$ statistical error bands. 
Right panels: Comparison of the same parton densities at different
orders in QCD as resulting from the present analysis.
}}
\end{figure}
%\normalsize
%%%%%%%%%%%%%%%%%%%%%%%%%%%%%%%%%%%%%%%%%%%%%%%%%%%%%%%%%%%%%%%%%%%%%%%
%%%%%%%%%%%%%%%%%%%%%%%%%%%%%%%%%%%%%%%%%%%%%%%%%%%%%%%%%%%%%%%%%%%%%%%
\newpage
%Figure~11:~
%\mbox{\epsfig{file=pdf_evol_nnlo_uv.eps,width=15cm}}
%\vspace{2mm}
%\noindent     
%\small   
%Figure~4:~
\begin{figure}
\includegraphics[angle=0, width=15.0cm]{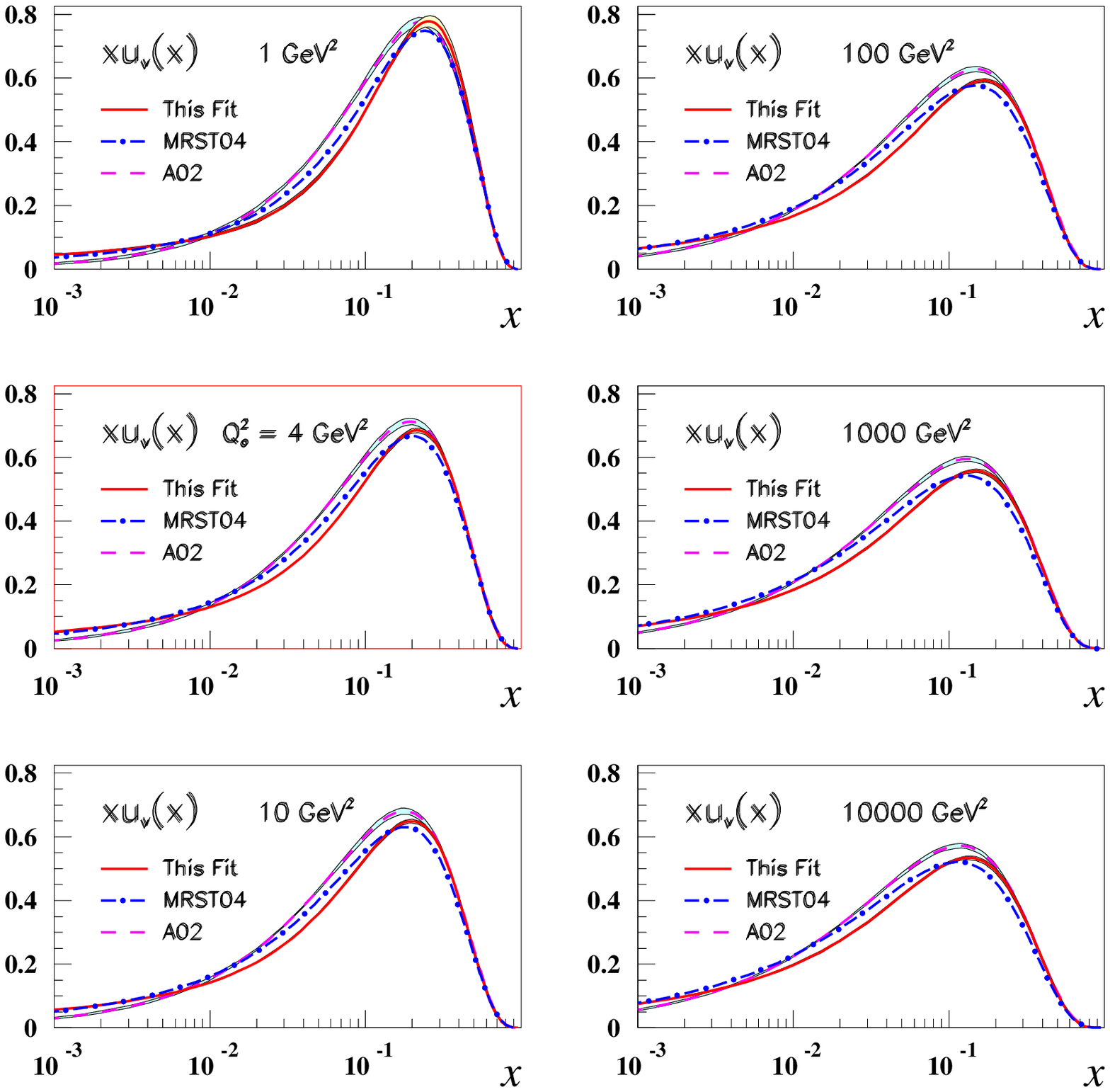} 
\begin{center}
\end{center}
{\sf
\caption{\label{fig:pdf_uv}
The parton density $xu_v$ at NNLO evolved up to $Q^2 = 10,000~{ \GeV^2}$
(solid lines) compared to results obtained by MRST (dashed--dotted
line)~\cite{MRST04} and 
A02 (dashed line)~\cite{A02}. The shaded
areas represent the fully correlated $1\sigma$ statistical error
bands.}}
\end{figure} 
%\normalsize
%%%%%%%%%%%%%%%%%%%%%%%%%%%%%%%%%%%%%%%%%%%%%%%%%%%%%%%%%%%%%%%%%%%%%%%
%%%%%%%%%%%%%%%%%%%%%%%%%%%%%%%%%%%%%%%%%%%%%%%%%%%%%%%%%%%%%%%%%%%%%%%
\newpage
%Figure~12:~
%\mbox{\epsfig{file=pdf_evol_nnlo_dv.eps,width=15cm}}
%\vspace{2mm}
%\noindent     
%\small   
%Figure~5:~
\begin{figure}
\includegraphics[angle=0, width=15.0cm]{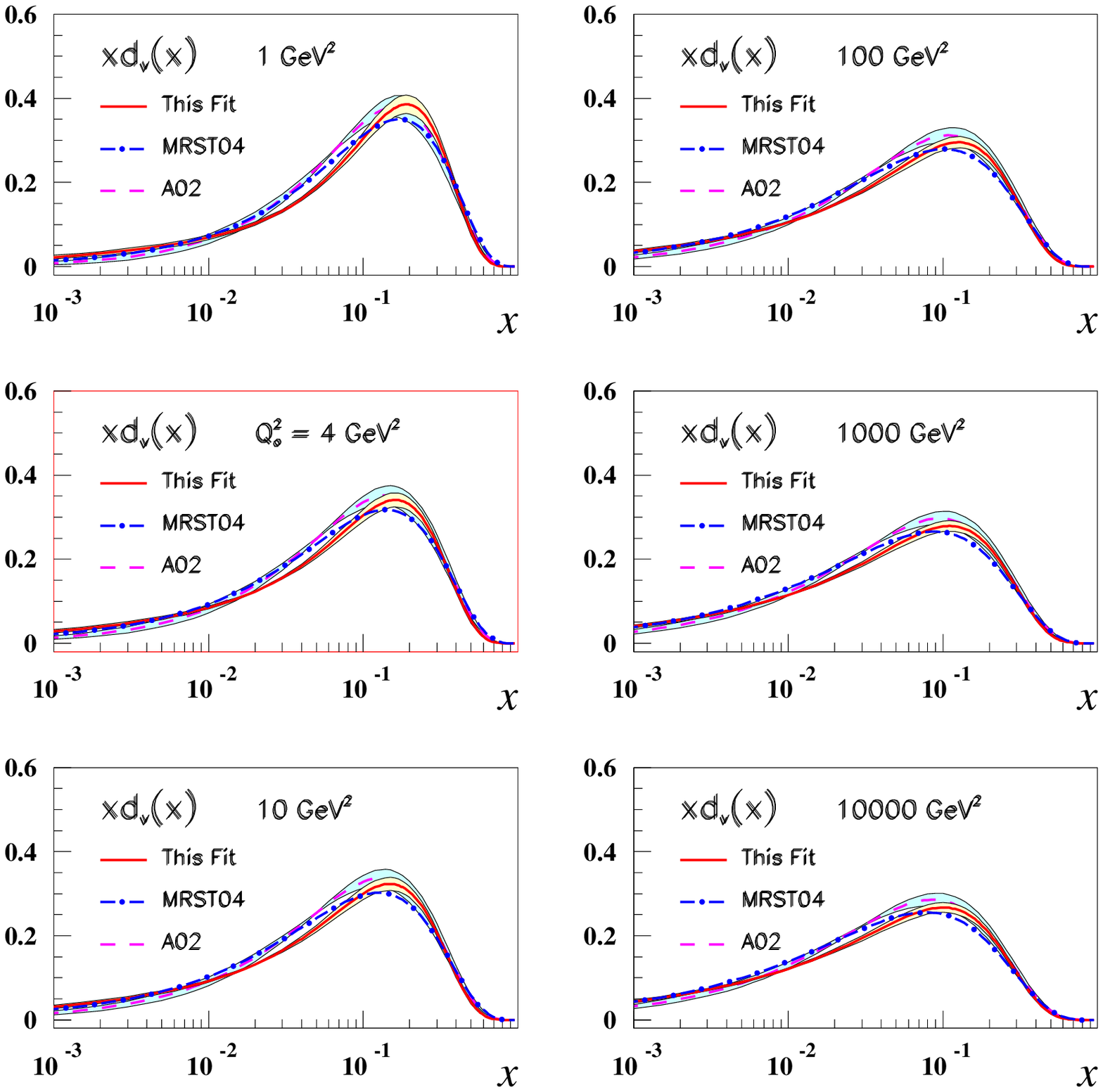} 
\begin{center}
\end{center}
{\sf
\caption{\label{fig:pdf_dv}
The parton density $xd_v$ at NNLO evolved up to $Q^2 = 10,000~{\GeV^2}$
(solid lines) compared to results obtained by MRST (dashed--dotted
line)~\cite{MRST04} and A02 (dashed line)~\cite{A02}. The shaded
areas represent the fully correlated $1\sigma$ statistical error
bands.}}
\end{figure} 
%\normalsize
%%%%%%%%%%%%%%%%%%%%%%%%%%%%%%%%%%%%%%%%%%%%%%%%%%%%%%%%%%%%%%%%%%%%%%%
%%%%%%%%%%%%%%%%%%%%%%%%%%%%%%%%%%%%%%%%%%%%%%%%%%%%%%%%%%%%%%%%%%%%%%%
\newpage
%Figure~13:~
%\mbox{\epsfig{file=HTCOEF_1_p.eps,width=15cm}}
%\vspace{2mm}
%\noindent     
%\small   
\begin{figure}
\begin{center}
\includegraphics[angle=0, width=15.0cm]{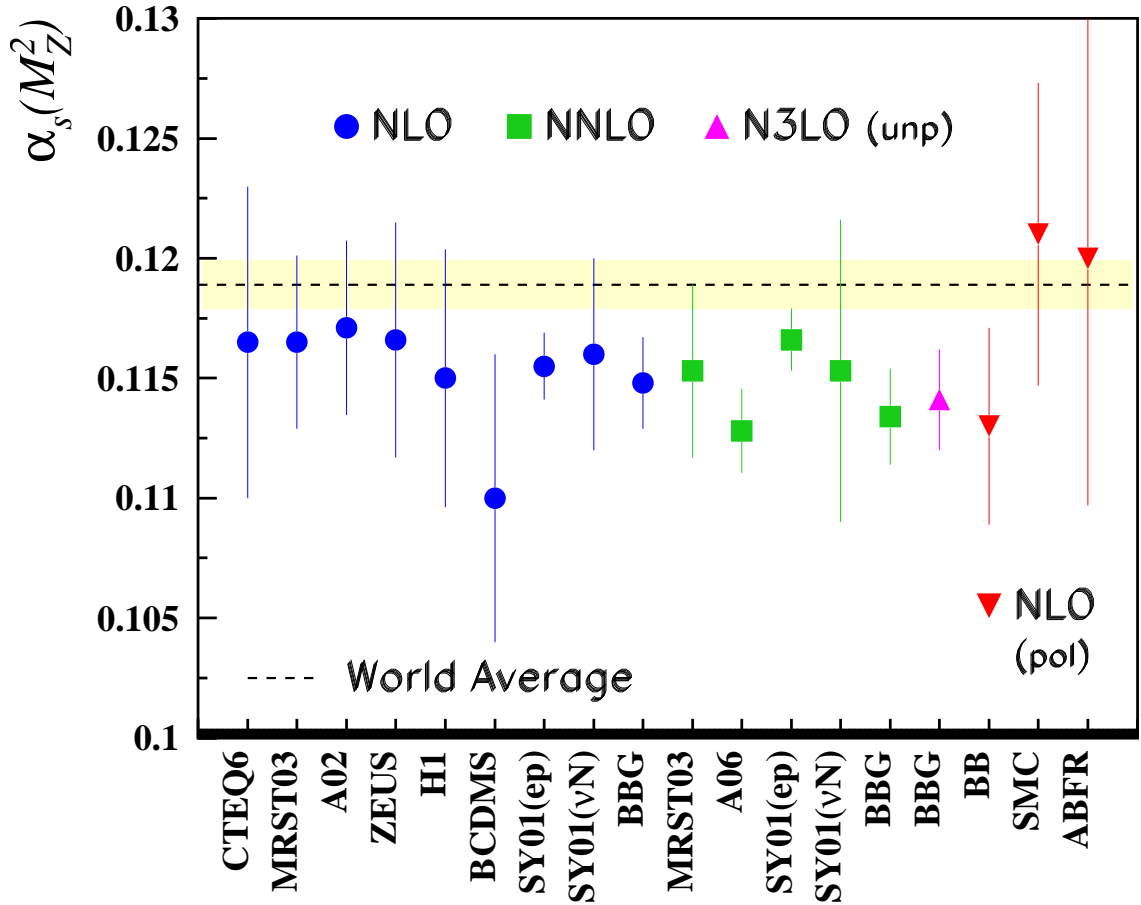} 
\end{center}
{\sf
\caption{\label{fig:alps}
Comparison of different NLO, NNLO and N$^3$LO measurements of the strong 
coupling constant $\alpha_s(M_Z^2)$ in unpolarized and polarized deeply inealstic 
scattering. 
CTEQ6 \cite{CTEQ_P}, MRST03 \cite{MRST03}, A02 \cite{A02},  ZEUS \cite{ZEUS_Ch}, 
H1 \cite{H1}, BCDMS \cite{BCDMS}, BBG: present analysis, 
SY01 \cite{SY}, 
A06 \cite{Alekhin:2006zm},
BB \cite{BB02}, 
SMC \cite{SMC}, ABFR \cite{ABFR}.}}
\end{figure}
\normalsize
%%%%%%%%%%%%%%%%%%%%%%%%%%%%%%%%%%%%%%%%%%%%%%%%%%%%%%%%%%%%%%%%%%%%%%%%
%%%%%%%%%%%%%%%%%%%%%%%%%%%%%%%%%%%%%%%%%%%%%%%%%%%%%%%%%%%%%%%%%%%%%%%
\newpage
%Figure~14:~
%\mbox{\epsfig{file=HTCOEF_1_p.eps,width=15cm}}
%\vspace{2mm}
%\noindent     
%\small   
\begin{figure}
\begin{center}
\includegraphics[angle=0, width=15.0cm]{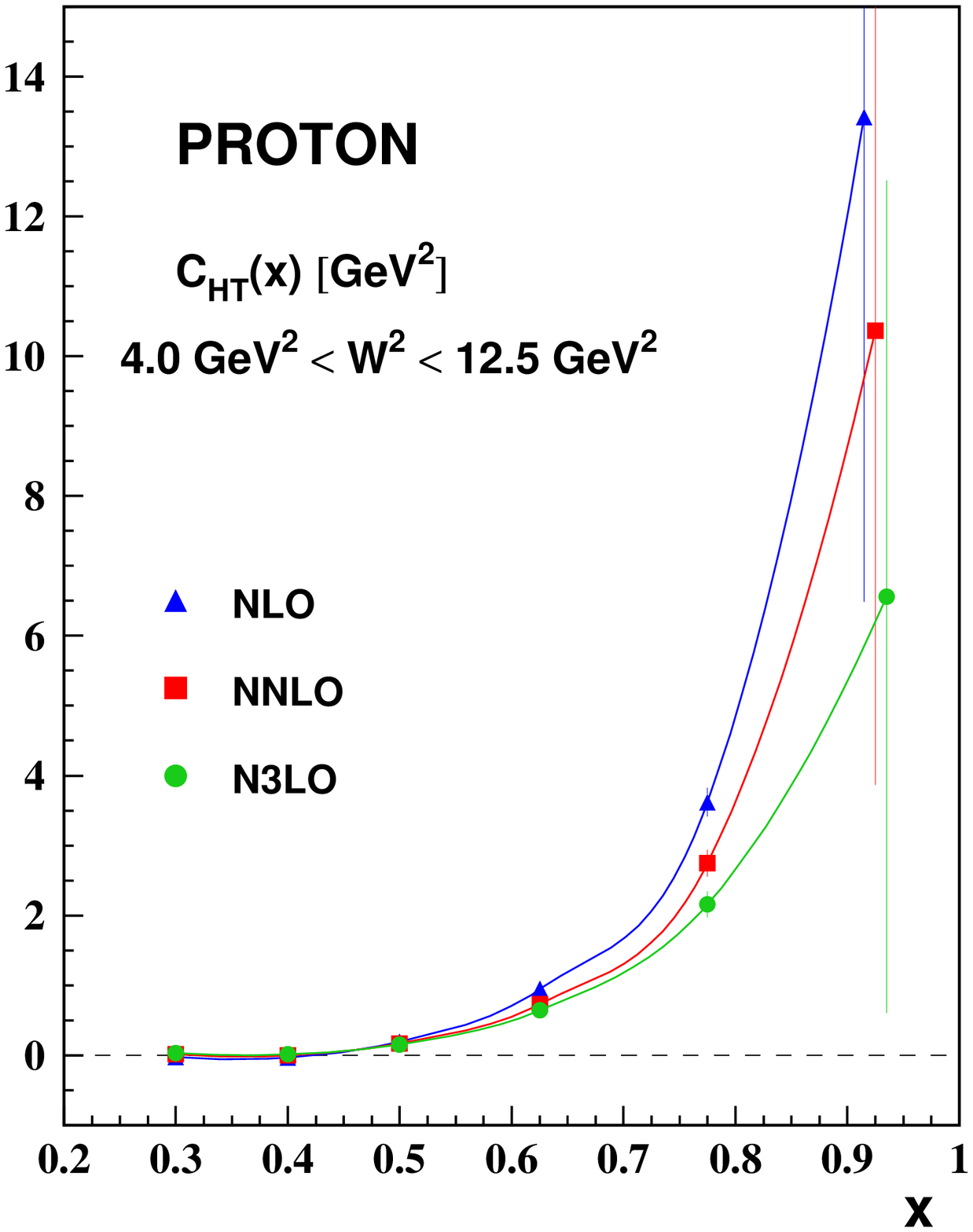} 
\end{center}
{\sf
\caption{\label{fig:HT_p}
The higher twist coefficient $C_{HT}(x)$ for the proton data as
function of $x$ treating the twist--2 contributions in NLO, NNLO and N$^3$LO.}}
\end{figure}
\normalsize
%%%%%%%%%%%%%%%%%%%%%%%%%%%%%%%%%%%%%%%%%%%%%%%%%%%%%%%%%%%%%%%%%%%%%%%%
%%%%%%%%%%%%%%%%%%%%%%%%%%%%%%%%%%%%%%%%%%%%%%%%%%%%%%%%%%%%%%%%%%%%%%%%
\newpage
%Figure~15:~
%\mbox{\epsfig{file=HTCOEF_1_d.eps,width=15cm}}
%\vspace{2mm}
%\noindent     
%\small   
\begin{figure}
\begin{center}
\includegraphics[angle=0, width=15.0cm]{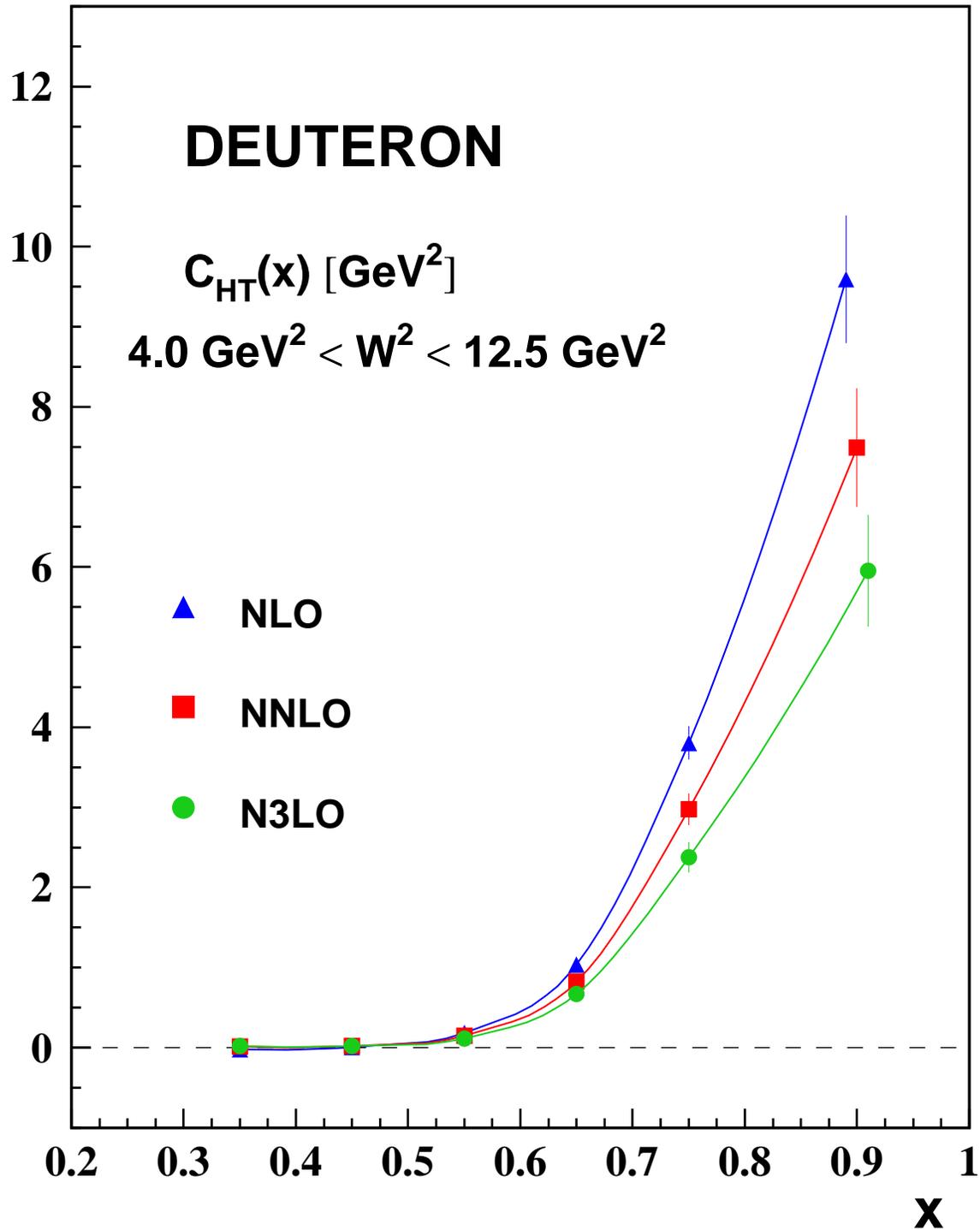} 
\end{center}
{\sf
\caption{\label{fig:HT_d}
The higher twist coefficient $C_{HT}(x)$ for the deuteron data as
function of $x$ treating the twist--2 contributions in NLO, NNLO and N$^3$LO.}}
\end{figure}
\normalsize
%%%%%%%%%%%%%%%%%%%%%%%%%%%%%%%%%%%%%%%%%%%%%%%%%%%%%%%%%%%%%%%%%%%%%%%
%%%%%%%%%%%%%%%%%%%%%%%%%%%%%%%%%%%%%%%%%%%%%%%%%%%%%%%%%%%%%%%%%%%%%%%
%	Bibliography
%%%%%%%%%%%%%%%%%%%%%%%%%%%%%%%%%%%%%%%%%%%%%%%%%%%%%%%%%%%%%%%%%%%%%%%
\clearpage
\portrait

%----------------------------------------------------------------------
%%%%%%%%%%%%%%%%%%%%%%%%%%%%%%%%%%%%%%%%%%%%%%%%%%%%%%%%%%%%%%%%%%%%%%%
%%%%%%%%%%%%%%%%%%%%%%%%%%%%%%%%%%%%%%%%%%%%%%%%%%%%%%%%%%%%%%%%%%%%%%%

\begin{thebibliography}{99}
%
%%%%%%%%%%%%%%%%%%%%%%%%%%%%%%%%%%%%%%%%%%%%%%%%%%%%%%%%%%%%%%%%%%%%%%%
%
%[1]
\bibitem{LC}
K.G.~Wilson, Phys. Rev. {\bf 179} (1969) 1699;\\
R.A.~Brandt and G.~Preparata, Fortschr. Phys. {\bf 18} (1970)
249;\\
W.~Zimmermann, {\sf Lect. on Elementary Particle Physics and
Quantum
Field Theory}, Brandeis Summer Inst., Vol.~1,
(MIT Press, Cambridge, 1970),~p. 395;\\
Y.~Frishman, Ann. Phys. {\bf 66} (1971) 373.
%------------------------------------------------------------------------
%
%[2]
\bibitem{BCDMS}
%\cite{Benvenuti:1989fm}
%\bibit{\it et al.}em{Benvenuti:1989fm}
A.C.~Benvenuti {\it et al.} [BCDMS Collaboration],
%``A High Statistics Measurement Of The Deuteron Structure Functions F2 (X,
%Q**2) And R From Deep Inelastic Muon Scattering At High Q**2,''
Phys.\ Lett.\ B {\bf 237} (1990) 592;
%%CITATION = PHLTA,B237,592;%%
\\
A.C.~Benvenuti {\it et al.} [BCDMS Collaboration], Phys. Lett. {\bf B223} (1989) 485;
Phys. Lett. {\bf B237} (1990) 592.
\\
%\cite{Benvenuti:1989gs}
%\bibitem{Benvenuti:1989gs}
A.C.~Benvenuti {\it et al.} [BCDMS Collaboration],
%``A Comparison Of The Structure Functions F2 Of The Proton And The Neutron From
%Deep Inelastic Muon Scattering At High Q**2,''
Phys.\ Lett.\ B {\bf 237} (1990) 599.
%%CITATION = PHLTA,B237,599;%%
%----------------------------------------------------------------------
%
%[3]
\bibitem{SLAC}
%\cite{Whitlow:1991uw}
%%\bibitem{Whitlow:1991uw}
L.~W.~Whitlow, E.~M.~Riordan, S.~Dasu, S.~Rock and A.~Bodek, %
% ``Precise measurements of the proton and deuteron structure functions from a
%global analysis of the SLAC deep inelastic electron scattering
%cross-sections,''
Phys.\ Lett.\ B {\bf 282} (1992). %
%%CITATION = PHLTA,B282,475;%%
%---------------------------------------------------------------------- 
%
%[4] 
\bibitem{NMC}
%\cite{Arneodo:1996qe}
%\bibitem{Arneodo:1996qe}
M.~Arneodo {\it et al.}  [New Muon Collaboration],
%``Measurement of the proton and deuteron structure functions, F2(p) and  F2(d),
%and of the ratio sigma(L)/sigma(T),''
Nucl.\ Phys.\ B {\bf 483} (1997) 3
[arXiv:hep-ph/9610231].
%%CITATION = HEP-PH 9610231;%%
%----------------------------------------------------------------------
%
%[5]
\bibitem{H1}
%\cite{Adloff:2000qk}
%\bibitem{Adloff:2000qk}
C.~Adloff {\it et al.}  [H1 Collaboration],
%``Deep-inelastic inclusive e p scattering at low x and a determination of
%alpha(s),''
Eur.\ Phys.\ J.\ C {\bf 21} (2001) 33
[arXiv:hep-ex/0012053];\\
%%CITATION = HEP-EX 0012053;%%
%\cite{Adloff:2003uh}
%\bibitem{Adloff:2003uh}
C.~Adloff {\it et al.}  [H1 Collaboration],
%``Measurement and QCD analysis of neutral and charged current cross  sections
%at HERA,''
Eur.\ Phys.\ J.\ C {\bf 30} (2003) 1
[arXiv:hep-ex/0304003].
%%CITATION = HEP-EX 0304003;%%
%----------------------------------------------------------------------
%
%[6]
\bibitem{ZEUS}
%\cite{Breitweg:1998dz}
%\bibitem{Breitweg:1998dz}
J.~Breitweg {\it et al.}  [ZEUS Collaboration],
%``ZEUS results on the measurement and phenomenology of F2 at low x and  low
%Q**2,''
Eur.\ Phys.\ J.\ C {\bf 7} (1999) 609
[arXiv:hep-ex/9809005];\\
%%CITATION = HEP-EX 9809005;%%
%\cite{Chekanov:2001qu}
%\bibitem{Chekanov:2001qu}
S.~Chekanov {\it et al.}  [ZEUS Collaboration],
%``Measurement of the neutral current cross section and F2 structure  function
%for deep inelastic e+ p scattering at HERA,''
Eur.\ Phys.\ J.\ C {\bf 21} (2001) 443
[arXiv:hep-ex/0105090].
%%CITATION = HEP-EX 0105090;%%
%----------------------------------------------------------------------
%
%[7]
%\cite{Moch:2004pa}
\bibitem{Moch:2004pa}
S.~Moch, J.~A.~M.~Vermaseren and A.~Vogt,
%``The three-loop splitting functions in QCD: The non-singlet case,''
Nucl.\ Phys.\ B {\bf 688} (2004) 101
[arXiv:hep-ph/0403192].
%%CITATION = HEP-PH 0403192;%%
%----------------------------------------------------------------------
%
%[8]
%\cite{Vermaseren:2005qc}
\bibitem{Vermaseren:2005qc}
J.~A.~M.~Vermaseren, A.~Vogt and S.~Moch,
%``The third-order QCD corrections to deep-inelastic scattering by photon
%exchange,''
Nucl.\ Phys.\ B {\bf 724} (2005) 3
[arXiv:hep-ph/0504242].
%%CITATION = HEP-PH 0504242;%%
%----------------------------------------------------------------------
%
%[9]
\bibitem{MOM}
%\cite{Larin:1993vu}
%\bibitem{Larin:1993vu}
S.~A.~Larin, T.~van Ritbergen and J.~A.~M.~Vermaseren,
%``The Next next-to-leading QCD approximation for nonsinglet moments of deep
%inelastic structure functions,''
Nucl.\ Phys.\ B {\bf 427} (1994) 41;\\
%%CITATION = NUPHA,B427,41;%%
%\cite{Larin:1996wd}
%\bibitem{Larin:1996wd}
S.~A.~Larin, P.~Nogueira, T.~van Ritbergen and J.~A.~M.~Vermaseren,
%``The 3-loop QCD calculation of the moments of deep inelastic  structure
%functions,''
Nucl.\ Phys.\ B {\bf 492} (1997) 338
[arXiv:hep-ph/9605317];\\
%%CITATION = HEP-PH 9605317;%%
%\cite{Retey:2000nq}
%\bibitem{Retey:2000nq}
A.~Retey and J.~A.~M.~Vermaseren,
%``Some higher moments of deep inelastic structure functions at  next-to-next-to
%leading order of perturbative QCD,''
Nucl.\ Phys.\ B {\bf 604} (2001) 281
[arXiv:hep-ph/0007294];\\
%%CITATION = HEP-PH 0007294;%%
%\cite{Blumlein:2004xt}
%\bibitem{Blumlein:2004xt}
J.~Bl\"umlein and J.~A.~M.~Vermaseren,
%``The 16th moment of the non-singlet structure functions F2(x,Q**2) and
%F(L)(x,Q**2) to O(alpha(s)**3),''
Phys.\ Lett.\ B {\bf 606} (2005) 130
[arXiv:hep-ph/0411111].
%%CITATION = HEP-PH 0411111;%%
%----------------------------------------------------------------------
%
%[10]
\bibitem{BC}
P.A. Baikov and K.G. Chetyrkin, in: Proceeedings of {\sf Loops and Legs in Quantum Field Theory, 2006},
Eisenach, April, 2006, Nucl. Phys. {\bf B} (Proc. Suppl.) to appear.
%----------------------------------------------------------------------
%
%[11]
\bibitem{MRST03}
%\cite{Martin:2003tt}
%\bibitem{Martin:2003tt}
A.~D.~Martin, R.~G.~Roberts, W.~J.~Stirling and R.~S.~Thorne,
%``MRST partons and uncertainties,''
arXiv:hep-ph/0307262.
%%CITATION = HEP-PH 0307262;%%
%----------------------------------------------------------------------
%
%[12]
%\cite{Alekhin:2002fv}
\bibitem{A02}
%\bibitem{Alekhin:2002fv}
S.~Alekhin,
%``Parton distributions from deep-inelastic scattering data,''
Phys.\ Rev.\ D {\bf 68} (2003) 014002
[arXiv:hep-ph/0211096].
%%CITATION = HEP-PH 0211096;%%
%----------------------------------------------------------------------
%
%[13]
%\cite{Alekhin:2006zm}
\bibitem{Alekhin:2006zm}
S.~Alekhin, K.~Melnikov and F.~Petriello,
%``Fixed target Drell-Yan data and NNLO QCD fits of parton distribution
%functions,''
arXiv:hep-ph/0606237 and private communication.
%%CITATION = HEP-PH 0606237;%%
%----------------------------------------------------------------------
%
%[14]
\bibitem{SY}
%\cite{Santiago:2001mh}
%\bibitem{Santiago:2001mh}
J.~Santiago and F.~J.~Yndurain,
%``Improved calculation of F2 in electroproduction and xF3 in neutrino
%scattering to NNLO and determination of alpha(s),''
Nucl.\ Phys.\ B {\bf 611} (2001) 447
[arXiv:hep-ph/0102247].
%%CITATION = HEP-PH 0102247;%%
%----------------------------------------------------------------------
%
%[15]
\bibitem{KAT}
%\cite{Kataev:2002rj}
%\bibitem{Kataev:2002rj}
A.~L.~Kataev, G.~Parente and A.~V.~Sidorov,
%``N**3LO fits to xF3 data: alpha(s) vs 1/Q**2 contributions,''
Nucl.\ Phys.\ Proc.\ Suppl.\  {\bf 116} (2003) 105
[arXiv:hep-ph/0211151];\\
%%CITATION = HEP-PH 0211151;%%
%\newpage
%\cite{Kataev:2001kk}
%\bibitem{Kataev:2001kk}
A.~L.~Kataev, G.~Parente and A.~V.~Sidorov,
%``Fixation of theoretical ambiguities in the improved fits to the xF3  CCFR
%data at the next-to-next-to-leading order and beyond,''
Phys.\ Part.\ Nucl.\  {\bf 34} (2003) 20
[Fiz.\ Elem.\ Chast.\ Atom.\ Yadra {\bf 34} (2003) 43]
[arXiv:hep-ph/0106221].
%%CITATION = HEP-PH 0106221;%%
%---------------------------------------------------------------------- 
%
%[16]
\bibitem{SYS}
%\cite{Santiago:1999pr}
%\bibitem{Santiago:1999pr}
J.~Santiago and F.~J.~Yndurain,
%``Calculation of electroproduction to NNLO and precision determination of
%alpha(s),''
Nucl.\ Phys.\ B {\bf 563} (1999) 45
[arXiv:hep-ph/9904344].
%%CITATION = HEP-PH 9904344;%%
%----------------------------------------------------------------------
% 
%[17] 
\bibitem{BBG04}
%\cite{Blumlein:2004ip}
%\bibitem{Blumlein:2004ip}
J.~Bl\"umlein, H.~B\"ottcher and A.~Guffanti,
%``Non-singlet QCD analysis of the structure function F2 in 3-loops,''
Nucl.\ Phys.\ Proc.\ Suppl.\  {\bf 135} (2004) 152
[arXiv:hep-ph/0407089];
%\cite{Blumlein:2006ws}
%\bibitem{Blumlein:2006ws}
%J.~Blumlein, H.~Bottcher and A.~Guffanti,
%``NNLO analysis of unpolarized DIS structure functions,''
arXiv:hep-ph/0606309.
%%CITATION = HEP-PH 0606309;%%
%----------------------------------------------------------------------
%
%[18]
%\cite{Gluck:2006yz}
\bibitem{Gluck:2006yz}
M.~Gl\"uck, E.~Reya and C.~Schuck,
%``Non-singlet QCD analysis of F2(x,Q**2) up to NNLO,''
arXiv:hep-ph/0604116.
%%CITATION = HEP-PH 0604116;%%
%----------------------------------------------------------------------
%
%[19]
\bibitem{CTEQ_P}
%\cite{Pumplin:2002vw}
%\bibitem{Pumplin:2002vw}
J.~Pumplin, D.~R.~Stump, J.~Huston, H.~L.~Lai, P.~Nadolsky and W.~K.~Tung,
%``New generation of parton distributions with uncertainties from global  QCD
%analysis,''
JHEP {\bf 0207} (2002) 012
[arXiv:hep-ph/0201195].
%%CITATION = HEP-PH 0201195;%%
%----------------------------------------------------------------------
%
%[20]
%\cite{Pumplin:2005rh}
\bibitem{Pumplin:2005rh}
J.~Pumplin, A.~Belyaev, J.~Huston, D.~Stump and W.~K.~Tung,
%``Parton distributions and the strong coupling: CTEQ6AB PDFs,''
JHEP {\bf 0602} (2006) 032
[arXiv:hep-ph/0512167].
%%CITATION = HEP-PH 0512167;%%
%----------------------------------------------------------------------
%
%[21]
\bibitem{FP}
%\cite{Furmanski:1981cw}
%\bibitem{Furmanski:1981cw}
W.~Furmanski and R.~Petronzio,
%``Lepton - Hadron Processes Beyond Leading Order In Quantum Chromodynamics,''
Z.\ Phys.\ C {\bf 11} (1982) 293.
%%CITATION = ZEPYA,C11,293;%%
%----------------------------------------------------------------------
%
%[22]
\bibitem{NS2}
%\cite{vanNeerven:1991nn}
%\bibitem{vanNeerven:1991nn}
W.~L.~van Neerven and E.~B.~Zijlstra,
%``Order alpha-s**2 contributions to the deep inelastic Wilson coefficient,''
Phys.\ Lett.\ B {\bf 272} (1991) 127;\\
%%CITATION = PHLTA,B272,127;%%
%\cite{Zijlstra:1992qd}
%\bibitem{Zijlstra:1992qd}
E.~B.~Zijlstra and W.~L.~van Neerven,
%``Order alpha-s**2 QCD corrections to the deep inelastic proton structure
%functions F2 and F(L),''
Nucl.\ Phys.\ B {\bf 383} (1992) 525.
%%CITATION = NUPHA,B383,525;%%
%----------------------------------------------------------------------
%
%[23]
%\cite{vanNeerven:1999ca}
\bibitem{vanNeerven:1999ca}
W.~L.~van Neerven and A.~Vogt,
%``NNLO evolution of deep-inelastic structure functions: The non-singlet
%case,''
Nucl.\ Phys.\ B {\bf 568} (2000) 263
[arXiv:hep-ph/9907472].
%%CITATION = HEP-PH 9907472;%%
%----------------------------------------------------------------------
%
%[24]
\bibitem{BK1}
%\cite{Blumlein:1998if}
%\bibitem{Blumlein:1998if}
J.~Bl\"umlein and S.~Kurth,
%``Harmonic sums and Mellin transforms up to two-loop order,''
Phys.\ Rev.\ D {\bf 60} (1999) 014018
[arXiv:hep-ph/9810241].
%%CITATION = HEP-PH 9810241;%%
%----------------------------------------------------------------------
%
%[25]
\bibitem{ANCONT}
%\cite{Blumlein:2000hw}
%bibitem{Blumlein:2000hw}
J.~Bl\"umlein,
%``Analytic continuation of Mellin transforms up to two-loop order,''
Comput.\ Phys.\ Commun.\  {\bf 133} (2000) 76
[arXiv:hep-ph/0003100].
%%CITATION = HEP-PH 0003100;%%
%----------------------------------------------------------------------
%
%[26]
\bibitem{JB1}
%\cite{Blumlein:2004bb}
%\bibitem{Blumlein:2004bb}
J.~Bl\"umlein,
%``Mathematical structure of anomalous dimensions and QCD Wilson  coefficients
%in higher order,''
Nucl.\ Phys.\ Proc.\ Suppl.\  {\bf 135} (2004) 225
[arXiv:hep-ph/0407044].
%%CITATION = HEP-PH 0407044;%%
%----------------------------------------------------------------------
%
%[27]
\bibitem{BM1}
J. Bl\"umlein  and S. Moch, to appear.
%----------------------------------------------------------------------
%
%[28]
\bibitem{BM2}
%\cite{Blumlein:2005jg}
%\bibitem{Blumlein:2005jg}
J.~Bl\"umlein and S.~O.~Moch,
%``Analytic continuation of the harmonic sums for the 3-loop anomalous
%dimensions,''
Phys.\ Lett.\ B {\bf 614} (2005) 53
[arXiv:hep-ph/0503188].
%%CITATION = HEP-PH 0503188;%%
%----------------------------------------------------------------------
%
%[29]
\bibitem{BR1}
%\cite{Blumlein:2005im}
%\bibitem{Blumlein:2005im}
J.~Bl\"umlein and V.~Ravindran,
%``Mellin moments of the next-to-next-to leading order coefficient  functions
%for the Drell-Yan process and hadronic Higgs-boson  production,''
Nucl.\ Phys.\ B {\bf 716} (2005) 128
[arXiv:hep-ph/0501178].
%%CITATION = HEP-PH 0501178;%%
%---------------------------------------------------------------------- 
% 
%[30] 
\bibitem{BR2}
%\cite{Blumlein:2006rr}
%\bibitem{Blumlein:2006rr}
J.~Bl\"umlein and V.~Ravindran,
%``O(alpha(s)**2) timelike Wilson coefficients for parton-fragmentation
%functions in Mellin space,''
Nucl.\ Phys.\ B {\bf 749} (2006) 1
[arXiv:hep-ph/0604019].
%%CITATION = HEP-PH 0604019;%%
%----------------------------------------------------------------------
%
%[31]
\bibitem{CHET}
%\cite{Chetyrkin:1997sg}
%\bibitem{Chetyrkin:1997sg}
K.~G.~Chetyrkin, B.~A.~Kniehl and M.~Steinhauser,
%``Strong coupling constant with flavour thresholds at four loops in the  MS-bar
%scheme,''
Phys.\ Rev.\ Lett.\  {\bf 79} (1997) 2184
[arXiv:hep-ph/9706430].
%%CITATION = HEP-PH 9706430;%%
%---------------------------------------------------------------------- 
% 
[32] 
\bibitem{SB}
%\cite{Bethke:2000ai}
%\bibitem{Bethke:2000ai}
S.~Bethke,
%``Determination of the QCD coupling alpha(s),''
J.\ Phys.\ G {\bf 26} (2000) R27 [arXiv:hep-ex/0004021].
%%CITATION = HEP-EX 0004021;%%
%----------------------------------------------------------------------
%
%[33]
\bibitem{BAR}
%\cite{Bardeen:1978yd}
%\bibitem{Bardeen:1978yd}
W.~A.~Bardeen, A.~J.~Buras, D.~W.~Duke and T.~Muta,
%``Deep Inelastic Scattering Beyond The Leading Order In Asymptotically Free
%Gauge Theories,''
Phys.\ Rev.\ D {\bf 18} (1978) 3998.
%%CITATION = PHRVA,D18,3998;%%
%----------------------------------------------------------------------
%
%[34]
\bibitem{R1998}
%\cite{Abe:1998ym}
%\bibitem{Abe:1998ym}
K.~Abe {\it et al.}  [E143 Collaboration],
%``Measurements of R = sigma(L)/sigma(T) for 0.03 < x < 0.1 and fit to  world
%data,''
Phys.\ Lett.\ B {\bf 452} (1999) 194
[arXiv:hep-ex/9808028].
%%CITATION = HEP-EX 9808028;%%
%---------------------------------------------------------------------- 
% 
%[35] 
\bibitem{MT}
%\cite{Melnitchouk:1995am}
%\bibitem{Melnitchouk:1995am}
W.~Melnitchouk and A.~W.~Thomas,
%``Q**2 dependence of nuclear shadowing,''
Phys.\ Rev.\ C {\bf 52} (1995) 3373
[arXiv:hep-ph/9508311].
%%CITATION = HEP-PH 9508311;%%
%----------------------------------------------------------------------
% 
%[36] 
\bibitem{JBMK}
J. Bl\"umlein and M. Klein, Nuclear Instruments and Methods in Physics 
Research, (NIM), {\bf A329} (1993) 112.
%----------------------------------------------------------------------
%
%[37]
\bibitem{dPDF}
C.Pascaud and E.Zomer, preprint LAL-95-05;\\ 
%\cite{Botje:1999dj}
%\bibitem{Botje:1999dj}
M.~Botje,
%``A QCD analysis of HERA and fixed target structure function data,''
Eur.\ Phys.\ J.\ C {\bf 14} (2000) 285
[arXiv:hep-ph/9912439];
%%CITATION = HEP-PH 9912439;%%
%\cite{Botje:2001fx}
%\bibitem{Botje:2001fx}
M.~Botje,
%``Error estimates on parton density distributions,''
J.\ Phys.\ G {\bf 28} (2002) 779
[arXiv:hep-ph/0110123;
%%CITATION = HEP-PH 0110123;%%
\\
%\cite{Stump:2001gu}
%\bibitem{Stump:2001gu}
D.~Stump {\it et al.},
%``Uncertainties of predictions from parton distribution functions. I:  The
%Lagrange multiplier method,''
Phys.\ Rev.\ D {\bf 65} (2002) 014012
[arXiv:hep-ph/0101051];
%%CITATION = HEP-PH 0101051;%%
\\
%\cite{Pumplin:2001ct}
%\bibitem{Pumplin:2001ct}
J.~Pumplin {\it et al.},
%``Uncertainties of predictions from parton distribution functions. II:  The
%Hessian method,''
Phys.\ Rev.\ D {\bf 65} (2002) 014013
[arXiv:hep-ph/0101032];
%%CITATION = HEP-PH 0101032;%%
\\
%\cite{Martin:2003sk}
%\bibitem{Martin:2003sk}
A.~D.~Martin, R.~G.~Roberts, W.~J.~Stirling and R.~S.~Thorne,
%``Uncertainties of predictions from parton distributions. II: Theoretical
%errors,''
Eur.\ Phys.\ J.\ C {\bf 35} (2004) 325;
[arXiv:hep-ph/0308087].
\\
%%CITATION = HEP-PH 0308087;%%
%\cite{Martin:2002aw}
%\bibitem{Martin:2002aw}
A.~D.~Martin, R.~G.~Roberts, W.~J.~Stirling and R.~S.~Thorne,
%``Uncertainties of predictions from parton distributions. I: Experimental
%errors. ((T)),''
Eur.\ Phys.\ J.\ C {\bf 28} (2003) 455
[arXiv:hep-ph/0211080].
%%CITATION = HEP-PH 0211080;%%
%----------------------------------------------------------------------
%
%[38]
\bibitem{MINUIT}
F.~James, CERN Program Library, Long Writeup D506 (MINUIT).
%----------------------------------------------------------------------
%
%[39]
\bibitem{E866}
%\cite{Towell:2001nh}
%\bibitem{Towell:2001nh}
R.~S.~Towell {\it et al.}  [FNAL E866/NuSea Collaboration],
%``Improved measurement of the anti-d/anti-u asymmetry in the nucleon sea,''
Phys.\ Rev.\ D {\bf 64} (2001) 052002
[arXiv:hep-ex/0103030].
%%CITATION = HEP-EX 0103030;%%
%----------------------------------------------------------------------
%
%[40]
\bibitem{MRST02}
%\cite{Martin:2001es}
%\bibitem{Martin:2001es}
A.~D.~Martin, R.~G.~Roberts, W.~J.~Stirling and R.~S.~Thorne,
%``MRST2001: Partons and alpha(s) from precise deep inelastic scattering  and
%Tevatron jet data,''
Eur.\ Phys.\ J.\ C {\bf 23} (2002) 73
[arXiv:hep-ph/0110215].
%----------------------------------------------------------------------
%
%[41]
\bibitem{GP76}
%\cite{Georgi:1976ve}
%\bibitem{Georgi:1976ve}
H.~Georgi and H.~D.~Politzer,
%``Freedom At Moderate Energies: Masses In Color Dynamics,''
Phys.\ Rev.\ D {\bf 14} (1976) 1829.
%%CITATION = PHRVA,D14,1829;%%
%%---------------------------------------------------------------------- 
%
%[42]
\bibitem{HF1}
E. Laenen, S. Riemersma, J. Smith, and W.L. van Neerven, Nucl. Phys. {\bf
B392} (1993) 162, 229;\\
S. Riemersma, J. Smith, and W.L. van Neerven,
Phys. Lett. {\bf B347} (1995) 143;\\
M. Buza, Y. Matiounine, J. Smith, R.L. Migneron and W.L. van Neerven,
Nucl. Phys. {\bf B472} (1996) 611.
%----------------------------------------------------------------------
%
%[43]
%\cite{Alekhin:2003ev}
\bibitem{Alekhin:2003ev}
S.~I.~Alekhin and J.~Bl\"umlein,
%``Mellin representation for the heavy flavor contributions to deep  inelastic
%structure functions,''
Phys.\ Lett.\ B {\bf 594} (2004) 299
[arXiv:hep-ph/0404034].
%%CITATION = HEP-PH 0404034;%%
%----------------------------------------------------------------------
%
%[44]
\bibitem{HF2}
J. Bl\"umlein, A. De Freitas, S. Klein, S. Moch and W.L. van Neerven, to appear.
%----------------------------------------------------------------------
%
%[45]
\bibitem{HF3}
%\cite{Blumlein:2006kc}
%\bibitem{Blumlein:2006kc}
J.~Blumlein,
%``O(alpha(s)**3) contributions to F(L)(Q anti-Q)(x, Q**2) for large
%virtualities,''
Nucl.\ Phys.\ Proc.\ Suppl.\  {\bf 157} (2006) 2.
%%CITATION = NUPHZ,157,2;%%
%----------------------------------------------------------------------
%
%[45]
%\cite{Kirschner:1983di}
\bibitem{Kirschner:1983di}
R.~Kirschner and L.~N.~Lipatov,
%``Double Logarithmic Asymptotics And Regge Singularities Of Quark Amplitudes
%With Flavor Exchange,''
Nucl.\ Phys.\ B {\bf 213} (1983) 122.
%%CITATION = NUPHA,B213,122;%%
%-------------------------------------------------------------------------------------------------
%
%[46]
\bibitem{BV}
%\cite{Blumlein:1995jp}
%\bibitem{Blumlein:1995jp}
J.~Bl\"umlein and A.~Vogt,
%``On the Behaviour of Non--Singlet Structure Functions at Small $x$,''
Phys.\ Lett.\ B {\bf 370} (1996) 149
[arXiv:hep-ph/9510410];
%%CITATION = HEP-PH 9510410;%%
%\cite{Blumlein:1996dd}
%\bibitem{Blumlein:1996dd}
%J.~Blumlein and A.~Vogt,
%``On the Resummation of $\alpha \ln~2 x$ Terms for Non-Singlet Structure
%Functions in QED and QCD,''
Acta Phys.\ Polon.\ B {\bf 27} (1996) 1309
[arXiv:hep-ph/9603450].
%%CITATION = HEP-PH 9603450;%%
%-------------------------------------------------------------------------------------------------
%
%[47]
\bibitem{BV1}
%\cite{Blumlein:1997em}
%\bibitem{Blumlein:1997em}
J.~Bl\"umlein and A.~Vogt,
%``The evolution of unpolarized singlet structure functions at small x,''
Phys.\ Rev.\ D {\bf 58} (1998) 014020
[arXiv:hep-ph/9712546];
%%CITATION = HEP-PH 9712546;%%
\\
%\cite{Blumlein:1996aw}
%\bibitem{Blumlein:1996aw}
J.~Bl\"umlein, S.~Riemersma and A.~Vogt,
%``The evolution of unpolarized and polarized structure functions at  small-x,''
Nucl.\ Phys.\ Proc.\ Suppl.\  {\bf 51C} (1996) 30
[arXiv:hep-ph/9608470];
%%CITATION = HEP-PH 9608470;%%
\\
%\cite{Blumlein:1998pp}
%\bibitem{Blumlein:1998pp}
J.~Bl\"umlein, V.~Ravindran, W.~L.~van Neerven and A.~Vogt,
%``The unpolarized gluon anomalous dimension at small x,''
arXiv:hep-ph/9806368.
%%CITATION = HEP-PH 9806368;%%
%\cite{Ball:1998be}
%\bibitem{Ball:1998be}
R.~D.~Ball and S.~Forte,
%``Corrections at small x,''
arXiv:hep-ph/9805315;\\
%%CITATION = HEP-PH 9805315;%%
%\cite{Blumlein:1999ev}
%\bibitem{Blumlein:1999ev}
J.~Bl\"umlein,
%``QCD evolution of structure functions at small x,''
arXiv:hep-ph/9909449.
%%CITATION = HEP-PH 9909449;%%
%-------------------------------------------------------------------------------------------------
%
%[48]
\bibitem{MRST04}
%\cite{Martin:2004ir}
%\bibitem{Martin:2004ir}
A.~D.~Martin, R.~G.~Roberts, W.~J.~Stirling and R.~S.~Thorne,
%``Physical gluons and high-E(T) jets,''
Phys.\ Lett.\ B {\bf 604} (2004) 61
[arXiv:hep-ph/0410230].
%%CITATION = HEP-PH 0410230;%%
%------------------------------------------------------------------------
%
%[49]
\bibitem{GRV98}
%\cite{Gluck:1998xa}
%\bibitem{Gluck:1998xa}
M.~Gl\"uck, E.~Reya and A.~Vogt,
%``Dynamical parton distributions revisited,''
Eur.\ Phys.\ J.\ C {\bf 5} (1998) 461
[arXiv:hep-ph/9806404].
%%CITATION = HEP-PH 9806404;%%
%------------------------------------------------------------------------
% 
%[63] 
\bibitem{PC}
S. Alekhin, private communication.
%----------------------------------------------------------------------
%
%[50]
\bibitem{KRI2}
V.G Krivokhizhin et al., Z. Phys. {\bf C48} (1990) 347.
%------------------------------------------------------------------------
%
%[51]
\bibitem{KRI3}
V.G Krivokhizhin and A.V. Kotikov, Physics of Atomic Nuclei,
{\bf 11} (2005) 1935.
%------------------------------------------------------------------------
%
%[53]
%\cite{Seligman:1997mc}
\bibitem{Seligman:1997mc}
W.~G.~Seligman {\it et al.},
%``Improved determination of alpha(s) from neutrino nucleon scattering,''
Phys.\ Rev.\ Lett.\  {\bf 79} (1997) 1213.
%%CITATION = PRLTA,79,1213;%%
%------------------------------------------------------------------------
%
%[54]
%\cite{Bethke:2006ac}
\bibitem{Bethke:2006ac}
S.~Bethke,
%``Experimental tests of asymptotic freedom,''
arXiv:hep-ex/0606035.
%%CITATION = HEP-EX 0606035;%%
%----------------------------------------------------------------------
%
%[55]
\bibitem{ZEUS_Ch}
%\cite{Chekanov:2002pv}
%\bibitem{Chekanov:2002pv}
S.~Chekanov {\it et al.}  [ZEUS Collaboration],
%``A ZEUS next-to-leading-order QCD analysis of data on deep inelastic
%scattering,''
Phys.\ Rev.\ D {\bf 67} (2003) 012007
[arXiv:hep-ex/0208023].
%%CITATION = HEP-EX 0208023;%%
%----------------------------------------------------------------------
%
%[56]
\bibitem{Thorne:2006zu}
R.~S.~Thorne, A.~D.~Martin and W.~J.~Stirling,
%``MRST Parton Distributions -- status 2006,''
arXiv:hep-ph/0606244.
%%CITATION = HEP-PH 0606244;%%
%----------------------------------------------------------------------
%
%[57]
\bibitem{BB02}
%\cite{Blumlein:2002be}
%\bibitem{Blumlein:2002be}
J.~Bl\"umlein and H.~B\"ottcher,
%``QCD analysis of polarized deep inelastic scattering data and parton
%distributions,''
Nucl.\ Phys.\ B {\bf 636} (2002) 225
[arXiv:hep-ph/0203155].
%%CITATION = HEP-PH 0203155;%%
%----------------------------------------------------------------------
%
%[58]
\bibitem{SMC}
%\cite{Adeva:1998vw}
%\bibitem{Adeva:1998vw}
B.~Adeva {\it et al.}  [Spin Muon Collaboration],
%``A next-to-leading order QCD analysis of the spin structure function g1,''
Phys.\ Rev.\ D {\bf 58} (1998) 112002.
%%CITATION = PHRVA,D58,112002;%%
%----------------------------------------------------------------------
%
%[59]
\bibitem{ABFR}
%\cite{Altarelli:1996nm}
%\bibitem{Altarelli:1996nm}
G.~Altarelli, R.~D.~Ball, S.~Forte and G.~Ridolfi,
%``Determination of the Bjorken sum and strong coupling from polarized
%structure functions,''
Nucl.\ Phys.\ B {\bf 496} (1997) 337
[arXiv:hep-ph/9701289].
%%CITATION = HEP-PH 9701289;%%
%----------------------------------------------------------------------
%
%[60]
%\cite{DellaMorte:2004bc}
\bibitem{DellaMorte:2004bc}
M.~Della Morte, R.~Frezzotti, J.~Heitger, J.~Rolf, R.~Sommer and U.~Wolff
                  [ALPHA Collaboration],
%``Computation of the strong coupling in QCD with two dynamical 
flavours,''
Nucl.\ Phys.\ B {\bf 713} (2005) 378
[arXiv:hep-lat/0411025].
%%CITATION = HEP-LAT 0411025;%%
%\cite{Gockeler:2005rv}
%----------------------------------------------------------------------
%
%[61]
\bibitem{Gockeler:2005rv}
M.~G\"ockeler, R.~Horsley, A.~C.~Irving, D.~Pleiter, P.~E.~L.~Rakow, 
G.~Schierholz and H.~Stuben,
%``A determination of the Lambda parameter from full lattice QCD,''
Phys.\ Rev.\ D {\bf 73} (2006) 014513
[arXiv:hep-ph/0502212].
%%CITATION = HEP-PH 0502212;%%
%----------------------------------------------------------------------
% 
%[62] 
\bibitem{VM92}
%\cite{Virchaux:1991jc}
%\bibitem{Virchaux:1991jc}
M.~Virchaux and A.~Milsztajn,
%``A Measurement of alpha-s and higher twists from a QCD analysis of high
%statistics F-2 data on hydrogen and deuterium targets,''
Phys.\ Lett.\ B {\bf 274} (1992) 221.
%%CITATION = PHLTA,B274,221;%%
%----------------------------------------------------------------------
\end{thebibliography}
\end{document}